\documentclass[longauth]{aa} 
\usepackage{graphicx}
\usepackage{natbib}
\usepackage{txfonts}
\bibpunct{(}{)}{;}{a}{}{,}

\def\deltaz{$3\le z \le 4$}
\def\z{\phantom{0}}
\def\dd{\mathrm{d}}
\begin{document}
   \title{The VIMOS VLT Deep Survey}
   \subtitle{The ultraviolet galaxy luminosity function and luminosity density at \deltaz\thanks{Based on observations
collected at the European Southern Observatory Very Large Telescope, Paranal, Chile, program 070.A-9007(A)}$^,$\thanks{Based on observations obtained at the Canada-France-Hawaii Telescope (CFHT), which is operated by the National Research Council of Canada, the Institut National des Sciences de l'Univers of the Centre National de la Recherche Scientifique of France,  and the University of Hawaii.}}

   \author{
     S. Paltani \inst{1,2,3}
\and O. Le F\`evre \inst{3}
\and O. Ilbert \inst{20}
\and S. Arnouts \inst{3}
\and S. Bardelli  \inst{5}
\and L. Tresse \inst{3}
\and G. Zamorani \inst{5} 
\and E. Zucca    \inst{5}
\and D. Bottini \inst{4}
\and B. Garilli \inst{4}
\and V. Le Brun \inst{3}
\and D. Maccagni \inst{4}
\and J.-P. Picat \inst{9}
\and R. Scaramella \inst{6,15}
\and M. Scodeggio \inst{4}
\and G. Vettolani \inst{6}
\and A. Zanichelli \inst{6}
\and C. Adami \inst{3}
\and M. Bolzonella  \inst{5} 
\and A. Cappi    \inst{5}
\and S. Charlot \inst{10,12}
\and P. Ciliegi    \inst{5}
\and T. Contini \inst{9}
\and S. Foucaud \inst{21}
\and P. Franzetti \inst{4}
\and I. Gavignaud \inst{14}
\and L. Guzzo \inst{11}
\and A. Iovino \inst{11}
\and H.J. McCracken \inst{12,13}
\and B. Marano     \inst{8}
\and C. Marinoni \inst{18}
\and A. Mazure \inst{3}
\and B. Meneux \inst{4,11}
\and R. Merighi   \inst{5} 
\and R. Pell\`o \inst{9}
\and A. Pollo \inst{3,17}
\and L. Pozzetti    \inst{5} 
\and M. Radovich \inst{7}
\and M. Bondi \inst{6}
\and A. Bongiorno \inst{8}
\and J. Brinchmann \inst{19}
\and O. Cucciati \inst{11,16}
\and S. de la Torre \inst{3}
\and F. Lamareille \inst{9}
\and Y. Mellier \inst{12,13}
\and P. Merluzzi \inst{7}
\and S. Temporin \inst{11}
\and D. Vergani \inst{4}
\and C.J. Walcher \inst{3}
}


   \institute{
Integral Science Data Centre, ch. d'\'Ecogia 16, CH-1290 Versoix, Switzerland\\
\email{Stephane.Paltani@obs.unige.ch}
\and
Geneva Observatory, University of Geneva, ch. des Maillettes 51, CH-1290 Sauverny, Switzerland
\and
Laboratoire d'Astrophysique de Marseille, UMR 6110 CNRS-Universit\'e de
Provence,  BP8, 13376 Marseille Cedex 12, France
\and
IASF-INAF - via Bassini 15, I-20133, Milano, Italy
\and
INAF-Osservatorio Astronomico di Bologna - Via Ranzani,1, I-40127, Bologna, Italy
\and
IRA-INAF - Via Gobetti,101, I-40129, Bologna, Italy
\and
INAF-Osservatorio Astronomico di Capodimonte - Via Moiariello 16, I-80131, Napoli,
Italy
\and
Universit\`a di Bologna, Dipartimento di Astronomia - Via Ranzani,1,
I-40127, Bologna, Italy
\and
Laboratoire d'Astrophysique de l'Observatoire Midi-Pyr\'en\'ees (UMR 
5572) -
14, avenue E. Belin, F31400 Toulouse, France
\and
Max Planck Institut fur Astrophysik, 85741, Garching, Germany
\and
INAF-Osservatorio Astronomico di Brera - Via Brera 28, Milan,
Italy
\and
Institut d'Astrophysique de Paris, UMR 7095, 98 bis Bvd Arago, 75014
Paris, France
\and
Observatoire de Paris, LERMA, 61 Avenue de l'Observatoire, 75014 Paris, 
France
\and
Astrophysical Institute Potsdam, An der Sternwarte 16, D-14482
Potsdam, Germany
\and
INAF-Osservatorio Astronomico di Roma - Via di Frascati 33,
I-00040, Monte Porzio Catone,
Italy
\and
Universit\'a di Milano-Bicocca, Dipartimento di Fisica - 
Piazza delle Scienze, 3, I-20126 Milano, Italy
\and
Astronomical Observatory of the Jagiellonian University, ul Orla 171, 
30-244 Krak{\'o}w, Poland
\and
Centre de Physique Th\'eorique, UMR 6207 CNRS-Universit\'e de Provence, 
F-13288 Marseille France
\and
Centro de Astrofísica da Universidade do Porto, Rua das Estrelas,
4150-762 Porto, Portugal 
\and
Institute for Astronomy, 2680 Woodlawn Dr., University of Hawaii,
Honolulu, Hawaii, 96822
\and	
 School of Physics \& Astronomy, University of Nottingham, University Park, Nottingham, NG72RD, UK
}

   \date{Received ; accepted }

 
  \abstract{
   {}
   {We study the luminosity function of the high-redshift galaxy population with redshifts \deltaz\ using a purely $I$-band magnitude-selected spectroscopic sample obtained in the framework of the VIMOS VLT Deep Survey (VVDS).}
   {We determine the luminosity function from the VVDS, taking care to add as few assumptions and as simple corrections as possible, and compare our results with those obtained from photometric studies, based on Lyman-break selections or photometric-redshift measurements.}
   {We find that in the redshift range \deltaz, the VVDS luminosity function is parameterized by $\phi^*=1.24^{+0.48}_{-0.50}\,10^{-3}$ mag$^{-1}$ Mpc$^{-3}$ and $M^*=-21.49^{+0.19}_{-0.19}$, assuming a slope $\alpha=-1.4$ consistent with most previous studies. While $\phi^*$ is comparable to previously found values, $M^*$ is significantly brighter by about 0.5\,mag at least. Using the conservative slope $\alpha=-1.4$, we find a luminosity density at 1700\,\AA\ ${\cal L}_{1700}(M<-18.5)=2.4\,10^{19}$ W Mpc$^{-3}$ and ${\cal L}_{1700}^{\mathrm{Total}}=3.1\,10^{19}$ W Mpc$^{-3}$, comparable to that estimated in other studies.}
   {The unexpectedly large number of very bright galaxies found in the VVDS indicates that the color-selection and photometric-redshift techniques that are generally used to build high-redshift galaxy samples may be affected by a significant fraction of color-measurement failures or by incomplete modelling of the mix of stellar emission, AGN contribution, dust absorption and intergalactic extinction assumed to identify high-redshift galaxies, making pure magnitude selection better able to trace the full population. Because of the difficulty to identify all low-luminosity galaxies in a spectroscopic survey, the luminosity density could still be significantly underestimated. We also find that the relative contribution of the most luminous galaxies compared to the fainter ones is at least twice as large in the VVDS compared to former estimates. Therefore, the VVDS paints a quite different picture of the role of the most actively star-forming galaxies in the history of star formation.}

}

\titlerunning{The VVDS UV galaxy LF and LD at \deltaz}

\authorrunning{S. Paltani et al.}

\maketitle

\section{Introduction}

Galaxies are formed and evolve through the influence of complex physical processes.  Galaxy number counts are one of the most obvious observational signatures of these processes, and constitute, as a consequence, a key ingredient to their understanding.  Notwithstanding how galaxies are assembled, the more galaxies there are, the more efficient star formation at earlier epochs must have been.  Counting the numbers of galaxies as a function of their total stellar luminosity and as a function of redshift in an exhaustive manner is therefore a necessary prerequisite to the determination of the total number of stars at a given epoch in the life of the universe.

Star-forming galaxies are most conspicuously revealed in the ultraviolet domain. The building of luminosity functions (LFs) in the ultraviolet, which describe the galaxy number densities as a function of their ultraviolet luminosities, is therefore an important step in the study of the star-formation history. At redshifts above two the detection of these galaxies is in theory made simpler by the fact that ultraviolet photons are redshifted in the optical range, which can be easily observed. However, building large samples of high-redshift galaxies is challenging.  At redshifts $z\gg 1$ the huge amount of observing time required to measure large numbers of redshifts has forced up to now to apply preselection criteria to weed out the much more numerous low-redshift galaxies.  A number of deep redshift surveys have identified various types of galaxies at redshifts above 2, including Lyman-break galaxies \citep[LBGs;][]{SteiEtal-1996-SpeCon}, distant red galaxies \citep{vanDEtal-2003-SpeCon}, BzK-selected galaxies \citep{DaddEtal-2004-NewPho}, faint K-selected galaxies \citep{AbraEtal-2004-GemDee}, or counterparts of faint sources detected in radio, X-ray or sub-mm surveys \citep{ChapEtal-2005-RedSur}.  In addition, a number of photometric-redshift surveys have identified high-redshift galaxies based on the expected spectrophotometric signature of combined stellar populations \citep[e.g.,][]{FontaEtal-2000-PhoRed,PoliEtal-2001-EvoLum,ArnoEtal-2002-MeaRed, Rowa-2003-PhoRed,GabaEtal-2004-EvoLum,SaraEtal-2006-ProEvo}. The understanding of the selection effects is however a serious challenge for the making of a complete census of the high-redshift galaxy population. In particular, the different criteria used to pick-up high-redshift galaxies have been shown to produce distinct, but overlapping populations \citep{ReddEtal-2005-CenOpt}.

In spite of these difficulties, several groups have built ultraviolet luminosity functions at redshift \deltaz. \citet{SteiEtal-1999-LymBre} built the first ultraviolet LFs at $z\sim 3$ and $z\sim 4$ based on the first large sample of high-redshift galaxies photometrically selected using the LBG color-color diagrams \citep{SteiEtal-1996-SpeCon}. Using a very similar approach, \citet{SawiThom-2005-UVGal} extended the study of the ultraviolet LF of LBGs using the small, but very deep, Keck Deep Field (KDF). At $z\sim 4$, \citet{OuchEtal-2004-CenLym} adapted the LBG selection technique to the filters used in the Subaru Deep Survey (SDS). Luminosity functions based on photometric-redshift studies have been calculated from samples built using deep imaging surveys, like the Hubble Deep Field \citep[HDF;][]{PoliEtal-2001-EvoLum,ArnoEtal-2005-MeaEvo} and the Fors Deep Field \citep[FDF;][]{GabaEtal-2004-EvoLum}. It must be pointed out that the difficulty in obtaining large samples without preselection have prevented up to now the study of high-redshift ultraviolet LFs based on spectroscopic samples, although spectroscopic confirmations have been obtained for the LBG sample of \citet{SteiEtal-1996-SpeCon}.

The VIMOS VLT Deep Survey \citep[VVDS;][]{LefeEtal-2005-FirEpo} is following a unique approach combining both a deep magnitude-selected spectroscopic survey and a large sample, an approach made possible by the recent appearance of the VIMOS high-multiplex multi-object spectrographs on the European Southern Observatory Very Large Telescope \citep{LefeEtal-2003-VirVLT}. The $I$-band selection of objects brighter than magnitude $I_{AB}=24$ allows us to identify galaxies up to $z\simeq 5$, and in the ``First Epoch'' VVDS we find 970 candidate galaxies with a redshift $1.4 \leq z \leq 5$ \citep{LefeEtal-2005-LarPop}. A surprising result is that the pure magnitude selection identifies an unexpectedly large population of galaxies at high redshift \citep{LefeEtal-2005-LarPop}, with surface densities of galaxies at redshifts $z\sim 2$ to $z\sim 4$ two to six times higher than reported by previous studies using LBG selection techniques.

In this Paper, we aim to quantify the luminosity function of the population of galaxies at high redshift found in the VVDS, concentrating on the \deltaz\ redshift range. We study the implications of the ultraviolet luminosity function on the ultraviolet luminosity density at these redshifts. The evolution of the ultraviolet luminosity density and the history of star formation are presented in \citet{TresEtal-2006-VVDSLD}.

Throughout the paper we use the concordance cosmology with $H_0=70$\,km s$^{-1}$ Mpc$^{-3}$, $\Omega_{\mathrm M}=0.3$ and $\Omega_\Lambda=0.7$.

\section{Data}

We use here the VVDS ``First Epoch'' sample in the VVDS-02h (+02h 26m -04$^\circ$ 30') field \citep{LefeEtal-2005-FirEpo}. This sample consists of the spectra of 9295 objects obtained over a sky area of 1720 arcmin$^2$ using the VIMOS multi-object spectrograph on the VLT unit 3 Melipal at ESO in Paranal \citep{LefeEtal-2003-VirVLT}.  Targets between limiting magnitudes $I_{\mathrm{AB}}=17.5$ and $I_{\mathrm{AB}}=24$ have been randomly selected.  On average, spectra have been obtained for 24\% of the sources in the photometric catalogue in the above magnitude range. The VIMOS observations have been carried out using the low-resolution red grism, which provides spectra with a resolution $\simeq 230$ over the 5500--9500\,\AA\ wavelength range. Typically 500-600 spectra are obtained simultaneously. The nominal exposure time is 16\,200\,s, split into 10 separate exposures of 27\,min, shifting the objects along the slits to reduce the effect of fringing at long wavelengths. Data reduction has been performed with the VIPGI data processing tool \citep{ScodEtal-2005-Vipgi}. Further details are given in \citet{LefeEtal-2005-FirEpo}.

As explained in \citet{LefeEtal-2005-FirEpo}, each spectrum has been assigned a flag which indicates the reliability of the redshift determination. Objects with flags 3 and 4 have very reliable measurements. A flag 2 indicates a reasonably reliable measurement, but with a non-negligible probability that the identified redshift is wrong; a flag 1 indicates a tentative redshift, with a significant probability of misidentification. When no redshift could be meaningfully estimated from a spectrum, the object gets a flag 0 and no redshift determination. In \citet{LefeEtal-2005-FirEpo}, confidence levels of 50\%, 80\%, 95\% and $\sim 100$\% have been estimated for the flag 1, 2, 3 and 4, respectively, on the basis of repeated observations; these levels have however been determined mostly for low-redshift galaxies, and \citet{LefeEtal-2005-LarPop} refined these estimates for specific redshift ranges at $z>1.4$. In total, among the 266 objects which have been attributed a redshift between 3 and 4, 12 objects have been attributed a flag 3 or 4, 101 a flag 2 and 153 a flag 1. In addition, 665 objects have a flag 0, and some of them may possibly be located at redshifts \deltaz. Objects classified as QSOs are excluded from our sample.

To build the luminosity functions, we complement our spectroscopic sample with multi-band photometry. The VVDS spectroscopic sample has been set up using photometric observations in the $B$, $V$, $R$ and $I$ bands obtained with the CFH12K CCD camera on the Canada-France-Hawaii telescope (CFHT) \citep{MccrEtal-2003-VdiII,LefeEtal-2004-VdiI}. In order to extend the photometric coverage down to the $U$ band, which is very important for the study of galaxies at $z\ge 3$, we use here deep $u^*$, $g'$, $r'$, $i'$ and $z'$ photometry obtained in the framework of the Legacy Survey project at the CFHT (CFHTLS\footnote{http://www.cfht.hawaii.edu/Science/CFHLS/}) with the MegaPrime/MegaCam imager.  Processing of the CFHTLS photometry is performed by the TERAPIX consortium\footnote{http://terapix.iap.fr}. CFHTLS's D1 deep field covers the VVDS-02h deep field. We use here the T0002 release, whose accumulated exposure times are, for the $u^*$, $g'$,$r'$, $i'$ and $z'$, 9.9\,hr, 5.4\,hr, 14.6\,hr, 33\,hr and 15\,hr respectively. The following 50\%-completeness limiting magnitudes have been obtained in the individual bands: $u^*_{50\%}=26.4$, $g'_{50\%}=26.3$, $r'_{50\%}=26.1$, $i'_{50\%}=25.9$ and $z'_{50\%}=24.9$. The CFHTLS photometric catalogue has been constructed using the VVDS astrometric reference frame, which shows excellent relative astrometric performances \citep{MccrEtal-2003-VdiII}. Therefore only a small matching aperture of $\simeq 0.3$ arcsec was needed when matching the VVDS spectroscopic catalog with the CFHTLS photometric catalog, avoiding multiple identifications.

\section{Weights and corrections in the calculation of the luminosity function}

While the VVDS selection function is very simple due to the VVDS being a purely magnitude-limited survey, the fraction of unobserved objects and errors in redshift determinations must be taken into account in the calculation of the luminosity functions. A specific weight is therefore applied to each object in order to recover the proper volume density at a given luminosity.  The weight includes three effects: the probability of a given object to have been observed spectroscopically; the probability of a given redshift to be correct; and the fraction of objects with a redshift in the range \deltaz\ mistakenly classified at other redshifts.

\subsection{Comparison with previous LF studies within the VVDS}
These difficulties have already been addressed in the framework of the VVDS for the determination of LFs at redshifts $z<2$ by \citet{IlbeEtal-2005-EvoGal}, and the methodology has been reused in subsequent works \citep{IlbeEtal-2006-EviEnv,IlbeEtal-2006-GalLum,ZuccEtal-LFTyp}. Except for the effect of probability of observation we use here a different approach from that used in these papers.

In \citet{IlbeEtal-2005-EvoGal}, flag-2, flag-3 and flag-4 objects have been assumed to have correct redshift determinations, while flag-1 objects have not been used. As already mentioned, the fraction of correct redshifts for flag-2 objects is about $80$\% at low redshift. The $\sim\!20$\% of wrong redshifts have however negligible impact on the determination of the LF, since, at low redshift, flag-2 objects constitute only about $40$\% of the objects with good-to-excellent redshift determination. At redshift \deltaz\ this figure reaches 90\%. Moreover, at these redshifts, the fraction of correct redshifts among the flag-2 objects is markedly lower than 80\%, and actually closer to $50$\% \citep{LefeEtal-2005-LarPop}. As a consequence, we cannot assume that all flag-2 objects have correct redshift determination, since the wrong-redshift objects would be almost as numerous as those with correct redshifts.

The discarding of all flag-1 objects in \citet{IlbeEtal-2005-EvoGal} makes that they must be treated as an incompleteness, similarly to the flag-0 objects. \citet{IlbeEtal-2005-EvoGal} summed the photometric-redshift probability densities for all flag-0 and flag-1 objects to determine the redshift distribution of these objects, which allowed them to estimate the fraction of missed objects (i.e., having received either a flag 0 or a flag 1) as a function of redshift and magnitude. Again, here we have to opt for a different approach, because photometric redshifts are not reliable enough at redshift $z>2$ to allow a proper estimate of the correction to be applied.

Below we explain in detail our treatment of the different effects.

\subsection{Sampling correction}
The \emph{target sampling rate} (TSR) is the probability that a given object in our field has been observed spectroscopically with VIMOS. The tool to place the slits for the VVDS Deep Field observations \citep[VMMPS;][]{BottEtal-2005-Vmmps} has been used in ``optimized mode'', which provides the largest multiplexing possible, instead of the ``random mode'', where all objects have the same probability to be observed. This has the consequence that the TSR of a given source depends on the size of the object, because extended objects have somewhat smaller chance of being targeted than point-like objects.  \citet{IlbeEtal-2005-EvoGal} estimated the TSR as a function of the X-radius parameter, which is the size of the object along the slit axis. We use here the same estimates, which are reported in figure 1 of \citet{IlbeEtal-2005-EvoGal}. To take this effect into account, the weight of source $i$ is multiplied by $1/{\mathit{TSR}}_i$. On average, the TSR is of the order of 24\%, and the effect of the X-radius parameter is small, the sources being mostly point-like.

\subsection{Wrongly assigned redshifts}
\label{sec:crf}
The galaxy counts must be corrected to account for the fraction of objects with wrong redshift determination. Following \citet{LefeEtal-2005-LarPop}, we try to estimate for each flag category the \emph{correct-redshift fraction} (CRF), i.e.\ the fraction of objects with correct redshift determination, by building the stacked composite spectrum of all flag-3/4 objects (whose CRFs are assumed to be equal to 1), which we compare with stacked composite spectra of flag-1 and flag-2 objects respectively. The equivalent widths of the different spectral lines are then compared to those found in the reference spectrum consisting of only flag-3/4 objects.  Assuming that the population of galaxies is similar for each flag, the average ratio between the equivalent widths in the less reliable-redshift stacked spectra and the reference spectrum provides an estimate of the CRF for the objects with quality flags 1 and 2.  We find a CRF of 0.44 for the flag-1 objects and of 0.49 for the flag-2 objects, consistent with the results of \citet{LefeEtal-2005-LarPop}, and again much lower than the value of 80\% found at lower redshift \citep{LefeEtal-2005-FirEpo}.

\begin{figure}[tb]
  \centering
  \includegraphics[width=8.8cm]{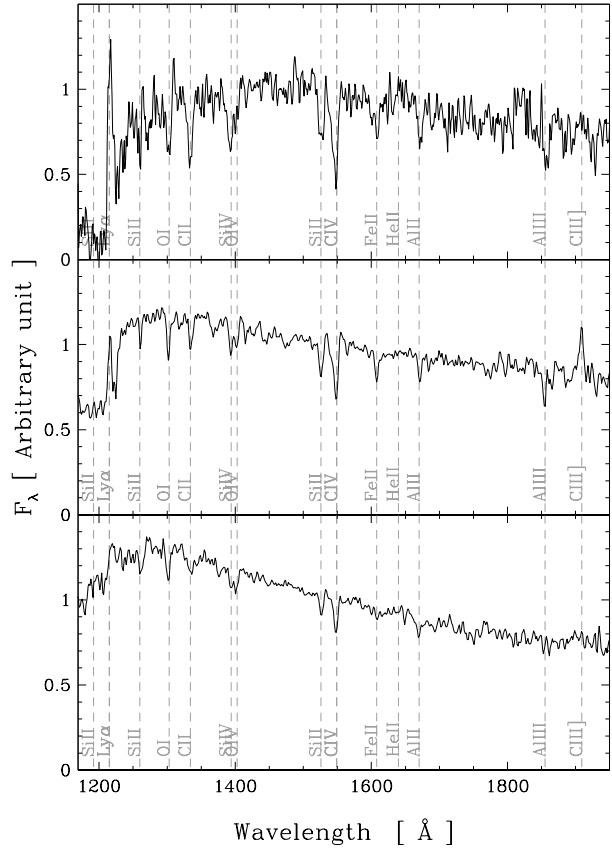}
  \caption{\label{fig:compo} Composite spectra of selected candidate \deltaz\ objects. Top panel: Composite spectrum of the 12 flag-3 and flag-4 objects. Middle panel: Composite spectrum of the 117 flag-1 and flag-2 spectra selected using the photometric-rejection method; flag-3 and flag-4 objects are not included. Bottom panel: Composite spectrum of the 137 flag-1 and flag-2 spectra rejected by the photometric-rejection method.  The major lines are indicated. The normalization has been chosen so that the equivalent widths of C\,{\sc iv} can be visually compared.}
\end{figure}

There are two ways to use the CRFs. The simplest assumption is that all sources in a flag category have the same probability to have a correct redshift. This corresponds to multiplying the weight of a given source with a flag $k$ by ${\mathit CRF_k}$, ${\mathit CRF_k}$ being equal to 0.44 for $k=1$ and to 0.49 for $k=2$; we refer to this method as the \emph{equal weight}.  Alternatively, as a redshift is either correct or wrong, we can try to identify objects that have smaller chances of having correct redshifts and discard them. The ``Algorithms for Luminosity Function'' \citep[ALF;][]{IlbeEtal-2005-EvoGal}, which is the tool developed in the framework of the VVDS consortium to calculate LFs, provides us with a quantitative criterion: When calculating absolute luminosities, ALF determines a merit figure based on the agreement, at the given redshift, between the best fitted template and the 9 photometric filters used here.  We can therefore, for flag-1 and flag-2 categories, select the ${\mathit CRF_k}\times N_k$ ($k=1,2$) objects having the best merit figures, where $N_k$ is the number of objects with flag $k$, and discard the remaining objects.  We refer to this method as the  \emph{photometric rejection}. We point out that mixing separate photometric sets can lead to problems with discrepant 0-points (in fact, this can happen even within a photometric system). \citet{IlbeEtal-2006-AccPho} determined that the most accurate photometric redshifts that can be obtained combining CFHTLS and CFH12K photometry requires adjustement in the 0-points of the different filters. However, these adjustments are always smaller than 0.1\,mag for the CWW templates \citep{ColeEtal-1980-ColMag}; we therefore neglect this correction here.

Fig.~\ref{fig:compo} shows the composite spectra of the 12 flag-3/4 objects, the 117 flag-1/2 selected spectra and of the 137 flag-1/2 rejected ones. The composite spectrum of the selected objects shows all the expected features of galaxies at $z\sim 3\!-\!4$, both in the spectral lines and in the continuum. While the composite of rejected spectra stills shows some of the lines expected in high-redshift galaxies (especially C\,{\sc iv}\,$\lambda 1549$\AA\ and Si\,{\sc ii}\,$\lambda 1527$\AA), they are significantly weaker and some strong lines, like Fe\,{\sc ii}\,$\lambda 1609$\AA, are missing. Part of the lines appearing in this composite might therefore be the result of wrong identification of features in the spectra with C\,{\sc iv} or Si\,{\sc ii} and possibly another shorter-wavelength line. In addition, the continuum shape is not really as expected, especially around Ly\,$\alpha$. We conclude that the photometric-rejection method very significantly increases the fraction of correctly-identified galaxies in our sample, and while it may be that some correctly-identified galaxies are improperly rejected, the remaining contamination is probably quite small.

\subsection{High-redshift galaxy incompleteness}
\label{sec:inc}
Incompleteness affects all types of surveys. For instance, to determine UV luminosity function at high redshift through LBG selection \citep[][ hereafter S99 and ST06A respectively]{SteiEtal-1999-LymBre,SawiThom-2005-UVGal}, the fraction of galaxies escaping the selection needs to be estimated using complex simulations of the effects of photometric incompleteness, unadapted flux integration aperture, contaminations, statistical scatter in the measured magnitudes, etc.  Thanks to the very simple selection function of the VVDS, such simulations, which rely on assumptions that are not always easy to assess, are not necessary. However, incompleteness does arise from the redshift-determination process through the lack of sufficient spectral information or the misidentification of the spectral features.

The redshift of some objects, especially the faintest ones, could not be determined; these objects are the so-called flag-0 objects, and contribute directly to the VVDS incompleteness. Assuming that the redshift distribution of these objects is not very different from that of the rest of the population, we can correct our galaxy counts to take these flag-0 objects into account: We estimate the fraction of flag-0 objects as a function of $I_{\mathrm{AB}}$ magnitude. This fraction, $f_0(I_{\mathrm{AB}})$, increases steeply at faint magnitudes. Thus, for each galaxy $i$ in the range \deltaz\ with a magnitude ${I_{\mathrm{AB}}}_i$, we apply a correction factor $1/(1-f_0({I_{\mathrm{AB}}}_i))$. This correction will have a negligible impact on the bright part of the LF, because the fraction of flag-0 objects is very low at bright magnitudes.

In \citet{IlbeEtal-2005-EvoGal} it has been shown that the redshift distribution of flag-2/4 objects is different from that of the flag-0/1 objects. If this is also the case here, the above correction may not be accurate. The situation is less clear in the redshift range \deltaz\, since the CRFs of both flag-1 and flag-2 objects are comparable to that of flag-1 objects at low redshift. In any case, this correction is probably valid only to the first order, and, in the following, we shall compare the LFs with and without this correction, which we call respectively ``corrected'' and ``minimal'' respectively.

Flag-1, and to a lesser extent flag-2, objects may have been wrongly assigned a redshift outside of the range \deltaz, and a full incompleteness correction should include a contribution from these flag-1 and flag-2 objects erroneously classified as low-redshift galaxies. However, such correction is more difficult to determine in a robust way, and we prefer not to take these objects into account, keeping in mind that the true galaxy counts are probably higher than those we calculated.

\subsection{Cosmic variance}
\begin{figure}[t]
   \centering
   \includegraphics[width=8.8cm]{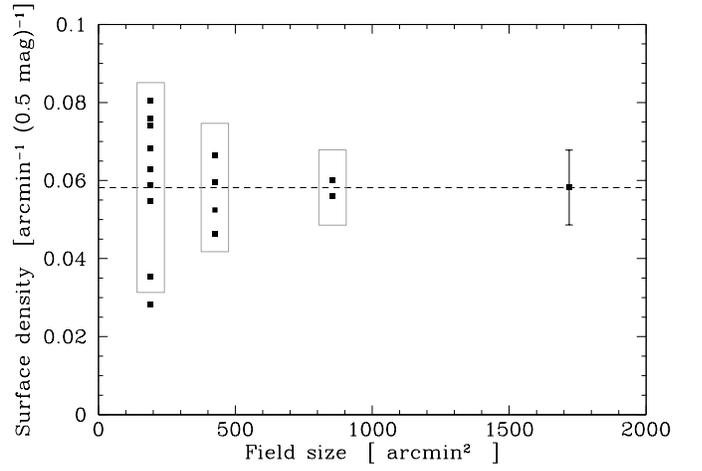}
   \caption{\label{fig:cosvar}Surface density of VVDS high-redshift objects in subfields of different sizes compared to that of the full field (from left to right, 3x3, 2x2, 1x2 and 1x1). The statistical uncertainy on the full field is indicated. The grey boxes show the expected statistical uncertainties for the smaller fields. Number counts are corrected for the estimated fraction of correct objects.}
 \end{figure}

Cosmic variance, i.e. the presence of inhomogeneities due to the large-scale structure of the Universe, can play a role in the estimation of observables like surface densities and LFs, in particular when one compares the results from different studies. The usual strategy to evaluate this effect is to use different fields and to analyze them separately (e.g., S99; ST06A).

The VVDS Deep Field consists of a single field, but we can have a grip on the amplitude of cosmic variance by dividing the field into smaller fields. Fig.~\ref{fig:cosvar} shows the surface density of the high-redshift galaxies discussed in \citet{LefeEtal-2005-LarPop} in the VVDS Deep Field, together with the surface densities calculated on sub-areas obtained by cutting the full field in 2, 4 and 9 smaller fields respectively. The purely statistical uncertainty on the full field can be extrapolated to the smaller fields by assigning the galaxies at random to the different subfields. Fig.~\ref{fig:cosvar} shows that the dispersion in the smaller fields is consistent with that expected from pure counting statistics on the galaxies in the full field, and therefore that any effect of cosmic variance in the VVDS field on these scales is quite small compared to the statistical uncertainties. This does not preclude that structures on scales significantly larger than the full VVDS field may exist; however, these structures are not predicted by cosmological simulations \citep{SomeEtal-2004-CosVar}. In the following, cosmic variance is assumed to be negligible and is therefore left out.

\section{Results}

\subsection{Ultraviolet luminosity function}

We calculate the luminosity function using ALF. ALF incorporates the classical estimator $1/V_{\mathrm{max}}$ \citep{Schm-1968-SpaDis}, the $C^+$ estimator \citep{ZuccEtal-1997-ESOSli}, the maximum-likelihood estimator SWML \citep{EfstEtal-1988-AnaCom} and the parametric maximum-likelihood estimator STY \citep{SandEtal-1979-VelFie}, which uses the distribution proposed by \citet{Sche-1976-AnaExp}:
\[
\phi(M) \dd M= 0.4\phi^*\ln(10)\left( 10^{0.4(M^*-M)} \right)^{\alpha+1}\!\exp\left({-10^{0.4(M^*-M)}}\right) \dd M
\]

In the discussion below, we shall focus on the standard parameters obtained from STY: the characteristic absolute magnitude $M^*$, the characteristic density $\phi^*$ (in mag$^{-1}$ Mpc$^{-3}$) and the slope $\alpha$.

\begin{figure*}[t]
  \centering
  \includegraphics[width=8.8cm]{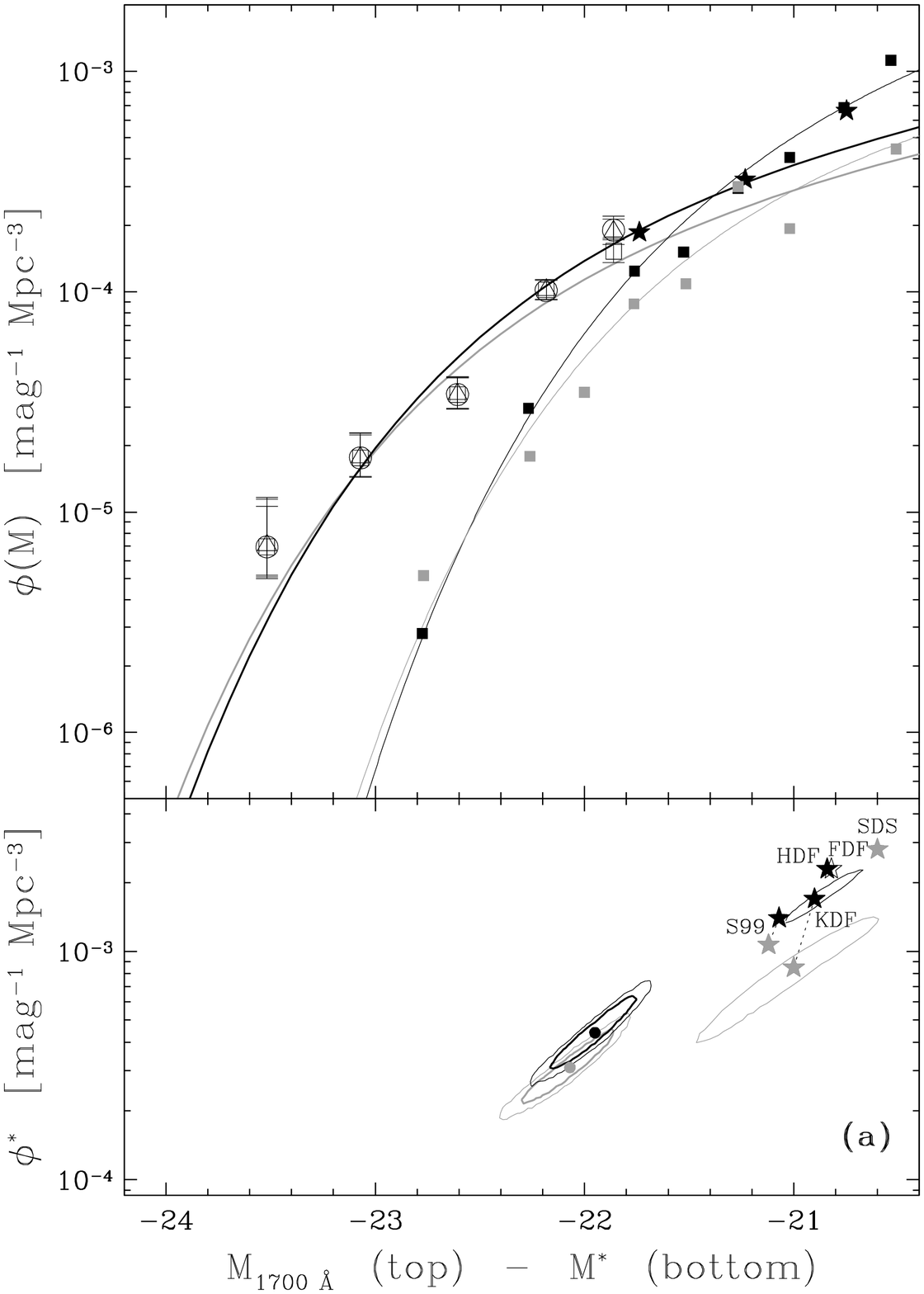}\hspace{3mm}\includegraphics[width=8.8cm]{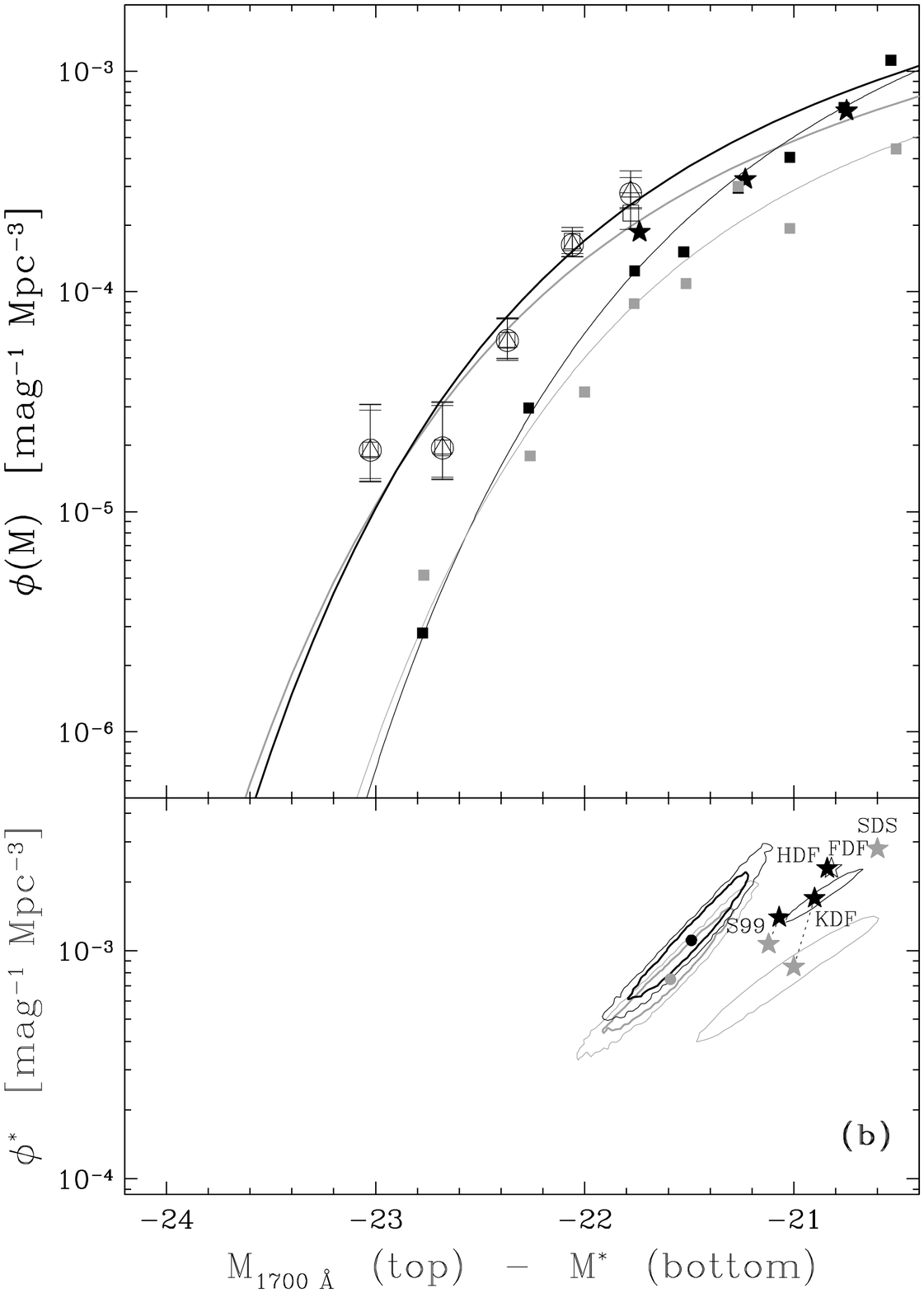}%
  \caption{\label{fig:LF} Luminosity functions for VVDS galaxies with \deltaz\ using the equal-weight method {\bf (a)} and the photometric-rejection method {\bf (b)}. Top panels: 1700\AA\ luminosity function as a function of M$_{1700}$. $\alpha$ has been fixed to -1.4. The heavy black and grey lines are the corrected and ``minimal'' LFs (see Sect.~\ref{sec:inc}), respectively. The empty symbols show the three non-parametric estimators for the corrected LF: $1/V_{\mathrm{max}}$ (squares), $C^+$ (triangles) and SWML (circles). The black and grey squares are the $1/V_{\mathrm{max}}$ estimators from ST06A at $z\sim 3$ and $z\sim 4$ respectively, and the thin lines are the corresponding STY LFs. The black stars are the non-parametric estimators of the $z\sim 3$ UV LF from \cite{PoliEtal-2001-EvoLum}. Bottom panel: Uncertainty contours of the 1700\,\AA\ luminosity function in the $M^*$--$\phi^*$ plane. The filled circles and the contours show the VVDS LFs; the minimal LF is shown in grey and the corrected LF in black. The heavy and thin solid lines are the $68$\% and $90$\% contours. Stars show the STY parameters obtained in other studies (see text); black, empty and grey stars are for $z\sim 3$, $z\sim 3.5$ and $z\sim 4$ respectively. The $68$\% uncertainty contours from ST06A are also shown.}
\end{figure*}

\begin{table}[t]
  \begin{center}
    \caption{Parameters of the UV luminosity functions estimated from the VVDS for different fixed values of $\alpha$. ``EW'' indicates the equal-weight LFs and ``PR'' the photometric-rejection ones. $\phi^*$ is in units of $10^{-3}$ mag$^{-1}$ Mpc$^{-3}$. The 68\% and 90\% (in parenthesis) confidence levels are also provided.}
    \label{tab:lfpar}
    \begin{tabular}{lccc}
      \hline
      \rule{0pt}{1.2em}%
      Method      &  $\phi^*$ & $M^*$ & $\alpha$ \\
      \hline
      \rule{0pt}{1.2em}%
                       &  $0.26^{+0.07}_{-0.07}\left(^{+0.13}_{-0.10}\right)$ & $-22.22^{+0.16}_{-0.17}\left(^{+0.26}_{-0.28}\right)$ & -1.6   \\
      ``EW'' minimal   &  $0.32^{+0.07}_{-0.08}\left(^{+0.14}_{-0.11}\right)$ & $-22.07^{+0.15}_{-0.15}\left(^{+0.23}_{-0.25}\right)$ & -1.4   \\
                       &  $0.38^{+0.08}_{-0.08}\left(^{+0.15}_{-0.12}\right)$ & $-21.89^{+0.13}_{-0.13}\left(^{+0.21}_{-0.22}\right)$ & -1.1   \\
      \rule{0pt}{1.5em}%
                       &  $0.38^{+0.10}_{-0.10}\left(^{+0.20}_{-0.15}\right)$ & $-22.08^{+0.15}_{-0.15}\left(^{+0.24}_{-0.26}\right)$ & -1.6   \\
      ``EW'' corrected &  $0.45^{+0.11}_{-0.11}\left(^{+0.21}_{-0.16}\right)$ & $-21.95^{+0.14}_{-0.14}\left(^{+0.22}_{-0.23}\right)$ & -1.4   \\
                       &  $0.53^{+0.11}_{-0.11}\left(^{+0.22}_{-0.17}\right)$ & $-21.78^{+0.12}_{-0.12}\left(^{+0.20}_{-0.21}\right)$ & -1.1   \\
      \rule{0pt}{1.5em}%
                       &  $0.75^{+0.32}_{-0.33}\left(^{+0.72}_{-0.44}\right)$ & $-21.71^{+0.22}_{-0.22}\left(^{+0.35}_{-0.38}\right)$ & -1.6   \\
      ``PR'' minimal &  $0.85^{+0.34}_{-0.34}\left(^{+0.73}_{-0.46}\right)$ & $-21.59^{+0.20}_{-0.21}\left(^{+0.33}_{-0.35}\right)$ & -1.4   \\
                       &  $0.94^{+0.33}_{-0.34}\left(^{+0.70}_{-0.46}\right)$ & $-21.44^{+0.18}_{-0.18}\left(^{+0.29}_{-0.30}\right)$ & -1.1   \\
      \rule{0pt}{1.5em}%
                       &  $1.14^{+0.49}_{-0.50}\left(^{+1.06}_{-0.66}\right)$ & $-21.59^{+0.20}_{-0.20}\left(^{+0.32}_{-0.35}\right)$ & -1.6   \\
      ``PR'' corrected &  $1.24^{+0.48}_{-0.50}\left(^{+1.06}_{-0.67}\right)$ & $-21.49^{+0.19}_{-0.19}\left(^{+0.30}_{-0.32}\right)$ & -1.4   \\
                       &  $1.34^{+0.48}_{-0.49}\left(^{+1.02}_{-0.66}\right)$ & $-21.35^{+0.17}_{-0.17}\left(^{+0.27}_{-0.29}\right)$ & -1.1   \\
      \hline
    \end{tabular}
  \end{center}
\end{table}

To allow comparisons with previous studies (e.g., S99, ST06A), we calculate the luminosity functions in a ultraviolet pseudo-filter centered around 1700\,\AA. The pseudo-filter we used is a 100\,\AA-wide box centered on 1700\,\AA, but the results do not depend on the details of the filter.  The redshift range $3\le z \le 4$ is particularly adapted to our observational set-up. At these redshifts, 1700\,\AA\ falls very close to, or even inside, the I filter used to select our galaxies. Absolute luminosities, which are calculated by fitting theoretical templates from the PEGASE2 library \citep{FiocRocc-1997-PegUV} to the full set of the nine photometric filters from the CFH12K and CFHTLS observations, will therefore need very small k-corrections. In addition, the redshift distribution of the objects in the VVDS falls quickly below $z\sim 3$ \citep{LefeEtal-2005-FirEpo} as a result of the so-called ``redshift desert'', produced by the combination of the observed wavelength range with VIMOS and the paucity of spectral lines and features in the rest-frame UV 1800--3500\,\AA\ galaxy spectra. The spectroscopic nature of the VVDS also limits the depth of the survey above $z\sim 4$.

In the redshift range \deltaz\ the VVDS reaches absolute magnitudes M$_{1700}\le -21.5$, which makes the $\alpha$ parameter completely unconstrained, as this limiting magnitude is brighter than M$^*$ found in previous studies of the UV LF. Other attempts to derive the high-redshift UV LF found rather conflicting values of $\alpha$, ranging from an assumed value of $-1.6$ in S99 to $\sim -1.4$ in ST06A, and even to $\sim -1.07$ in the FORS Deep Field study \citep[FDF;][]{GabaEtal-2004-EvoLum}. Instead of trying to determine the value of $\alpha$, we fixed it to $-1.6$, $-1.4$ and $-1.1$ to evaluate its effect. As most recent studies found values compatible with $-1.4$ (see Sect.~\ref{sec:comp}), we shall consider this value of $\alpha$ more closely. The resulting equal-weight LFs with $\alpha=-1.4$ are presented in Fig.~\ref{fig:LF}a, and the photometric-rejection LFs in Fig.~\ref{fig:LF}b. The non-parametric estimators are quite consistent with each other even in the lowest-luminosity bin, which indicates that there is little bias in the population under study \citep{IlbeEtal-2004-BiaEst}. Uncertainty contours in the $M^*$--$\phi^*$ plane for $\alpha=-1.4$ are plotted in the bottom panels of Figs~\ref{fig:LF}a and \ref{fig:LF}b. The Schechter parameters from the STY estimator are listed in Table~\ref{tab:lfpar}.

The choice of the method (equal-weight vs photometric-rejection) has a significant impact on $M^*$. Other parameters being equal, its value is about $0.5$ magnitude brighter in the equal-weight LFs than in the photometric-rejection ones. This is consistent with the rejection by the latter method of low-redshift galaxies that were mistakenly classified as high-redshift galaxies. As the density drops rapidly below $M^*$, even a small number of these spurious galaxies may alter significantly the estimate of $M^*$. Therefore we consider the photometric-rejection LFs to be more robust and adopt the results from this method as our best estimates.

The correction for flag-0 objects increases $\phi^*$ by about 40-50\%, but also increases $M^*$ by about 0.1\,mag, both effects being expected, since the correction gives more weight to the faintest galaxies in our sample. This correction is quite conservative, because an additional contribution from flag-1 and flag-2 objects erroneously classified as low-redshift galaxies is to be expected in addition to the flag-0 objects (see Sect.~\ref{sec:inc}). However, this additional correction, which would also increase $\phi^*$, is probably small and less straightforward to include in a robust way, and we did not attempt to include it.

The error contours in Figs~\ref{fig:LF}a and \ref{fig:LF}b show that, for a fixed slope $\alpha=-1.4$, $M^*$ and $\phi^*$ are quite strongly correlated, a bright $M^*$ implying a low $\phi^*$ and vice versa. The same effect can be seen by varying $\alpha$, with a change on $M^*$ of about $\pm 0.15$\,mag between $\alpha=-1.1$ and $\alpha=-1.6$.

In conclusion, when fixing the slope to $\alpha=-1.4$, the best estimate of the UV 1700\,\AA\ LF in the range \deltaz\ using the VVDS spectroscopic survey has the following STY parameters: $\phi^*=1.24^{+0.48}_{-0.50}\,10^{-3}$ mag$^{-1}$ Mpc$^{-3}$ and $M^*=-21.49^{+0.19}_{-0.19}$ (68\% confidence levels).

\begin{table}[t]
  \begin{center}
    \caption{\label{tab:ld}UV luminosity density ${\cal L}_{1700}(M<-18.5)$ integrated up to $M_{1700}=-18.5$ and total ${\cal L}_{1700}^{\mathrm{Total}}$ estimated from the VVDS for different fixed values of $\alpha$. ``EW'' indicates the equal-weight LFs and ``PR'' the photometric-rejection ones. The 68\% and 90\% (in parenthesis) confidence levels are also provided. Units are $10^{18}$ W Mpc$^{-3}$.}
    \begin{tabular}{lccc}
      \hline
      \rule{0pt}{1.2em}%
      Method      &  ${\cal L}_{1700}(M<-18.5)$ &  ${\cal L}_{1700}^{\mathrm{Total}}$ & $\alpha$ \\
      \hline
      \rule{0pt}{1.2em}%
                       &  $13.88^{+1.55}_{-1.57}\left(^{+2.76}_{-2.37}\right)$ &  $  19.20^{+2.59}_{-2.50}\left(^{+4.47}_{-3.79}\right)$ & -1.6   \\
      ``EW'' minimal   &  $11.78^{+1.24}_{-1.26}\left(^{+2.25}_{-1.90}\right)$ &  $  13.91^{+1.60}_{-1.64}\left(^{+3.00}_{-2.48}\right)$ & -1.4   \\
                       &  $\z 9.65^{+0.95}_{-0.99}\left(^{+1.84}_{-1.47}\right)$ &  $ 10.29^{+1.07}_{-1.12}\left(^{+2.06}_{-1.64}\right)$ & -1.1   \\
      \rule{0pt}{1.5em}%
                       &  $17.54^{+2.17}_{-2.17}\left(^{+3.94}_{-3.38}\right)$ &  $  24.75^{+3.52}_{-3.50}\left(^{+6.45}_{-5.33}\right)$ & -1.6   \\
      ``EW'' corrected &  $14.84^{+1.68}_{-1.70}\left(^{+3.03}_{-2.67}\right)$ &  $  17.75^{+2.21}_{-2.22}\left(^{+4.09}_{-3.51}\right)$ & -1.4   \\
                       & $12.02^{+1.26}_{-1.28}\left(^{+2.33}_{-2.00}\right)$ &  $  12.90^{+1.41}_{-1.43}\left(^{+2.62}_{-2.25}\right)$ & -1.1   \\
      \rule{0pt}{1.5em}%
                       &  $22.15^{+5.03}_{-5.03}\left(^{+9.89}_{-6.88}\right)$ &  $  33.50^{+9.00}_{-8.72}\left(^{+17.5}_{-11.9}\right)$ & -1.6   \\
      ``PR'' minimal &  $18.45^{+3.88}_{-3.89}\left(^{+8.38}_{-5.61}\right)$ &  $  23.15^{+5.41}_{-5.34}\left(^{+11.7}_{-7.78}\right)$ & -1.4   \\
                       &  $14.64^{+2.92}_{-2.97}\left(^{+5.61}_{-4.27}\right)$ &  $  16.11^{+3.41}_{-3.45}\left(^{+6.61}_{-4.95}\right)$ & -1.1   \\
      \rule{0pt}{1.5em}%
                       &  $29.20^{+6.64}_{-6.97}\left(^{+13.2}_{-9.85}\right)$ &  $  45.07^{+11.9}_{-12.3}\left(^{+24.1}_{-17.1}\right)$ & -1.6   \\
      ``PR'' corrected &  $24.31^{+5.32}_{-5.49}\left(^{+11.0}_{-7.93}\right)$ &  $  30.96^{+7.54}_{-7.75}\left(^{+15.3}_{-11.1}\right)$ & -1.4   \\
                       &  $18.98^{+3.88}_{-3.91}\left(^{+7.64}_{-5.51}\right)$ &  $  21.04^{+4.56}_{-4.59}\left(^{+9.18}_{-6.44}\right)$ & -1.1   \\
      \hline
    \end{tabular}
  \end{center}
\end{table}

\subsection{Ultraviolet luminosity density}

The luminosity density (LD) at 1700\,\AA\ ${\cal L}_{1700}$, which is the total power per volume unit contained in 1700\,\AA\ rest-wavelength photons emitted by galaxies, can be calculated by summing, possibly up to a limiting magnitude $M_{\mathrm{lim}}$, the contributions of the individual galaxies weighted by their number density given by the LF:
\begin{equation}
{\cal L}_{1700}(M\leq M_{\mathrm{lim}}) = \int_{-\infty}^{M_{\mathrm{lim}}}\,L_{1700}(M)\,\phi(M) \dd M,
\label{eq:ld}
\end{equation}
where $L_{1700}(M_{1700})=10^{-0.4(M_{1700}-34.1)}$ W Hz$^{-1}$ is the luminosity at 1700\,\AA\ of a galaxy with an absolute magnitude $M_{1700}$. The luminosity density ${\cal L}$ is expressed in W Hz$^{-1}$ Mpc$^{-3}$; the total LD ${\cal L}_{1700}^{\mathrm{Total}}$ is the limit $M_{\mathrm{lim}}\rightarrow\infty$.

\begin{figure*}[t]
  \centering
  \includegraphics[width=8.8cm]{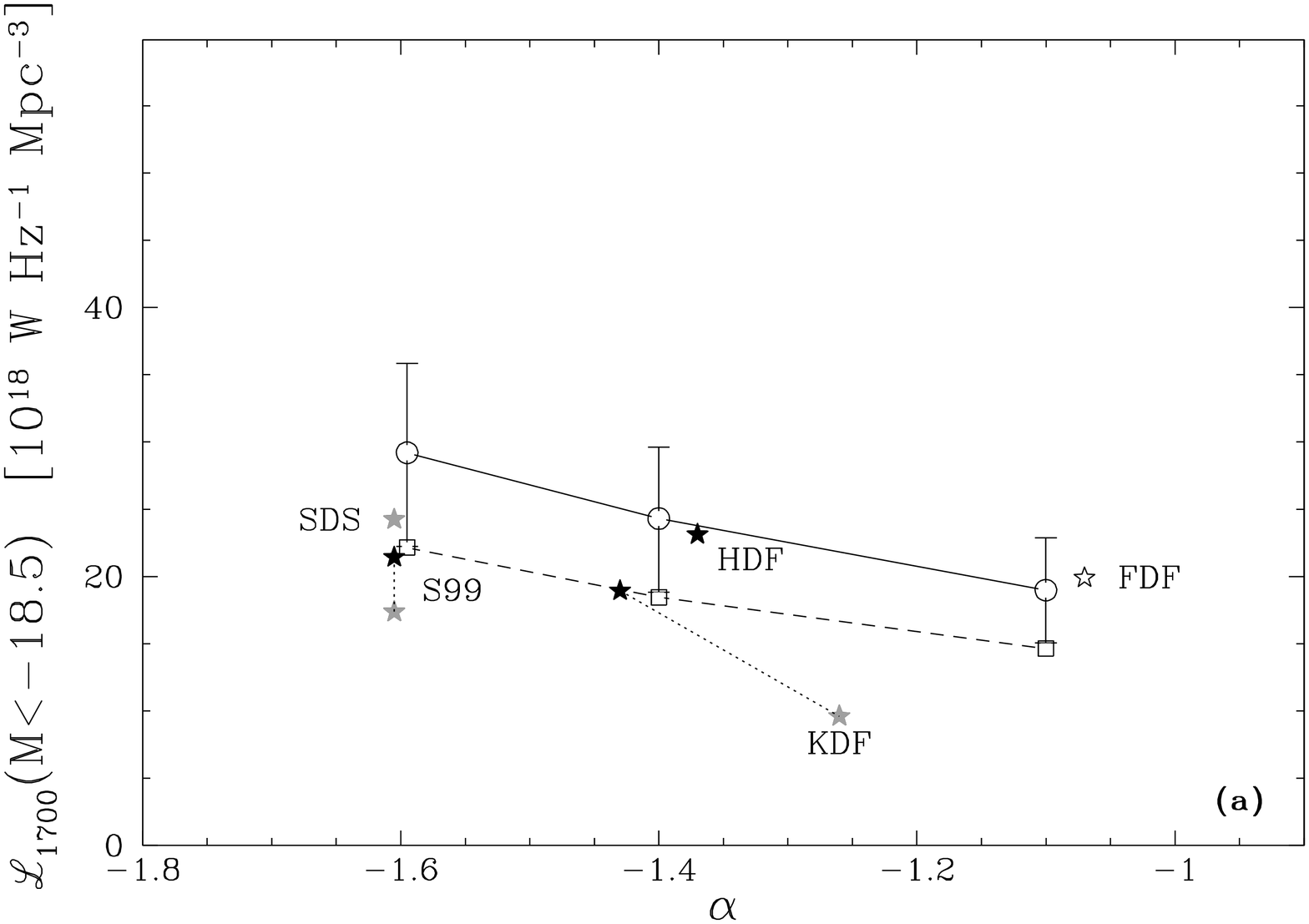}\hspace{3mm}\includegraphics[width=8.8cm]{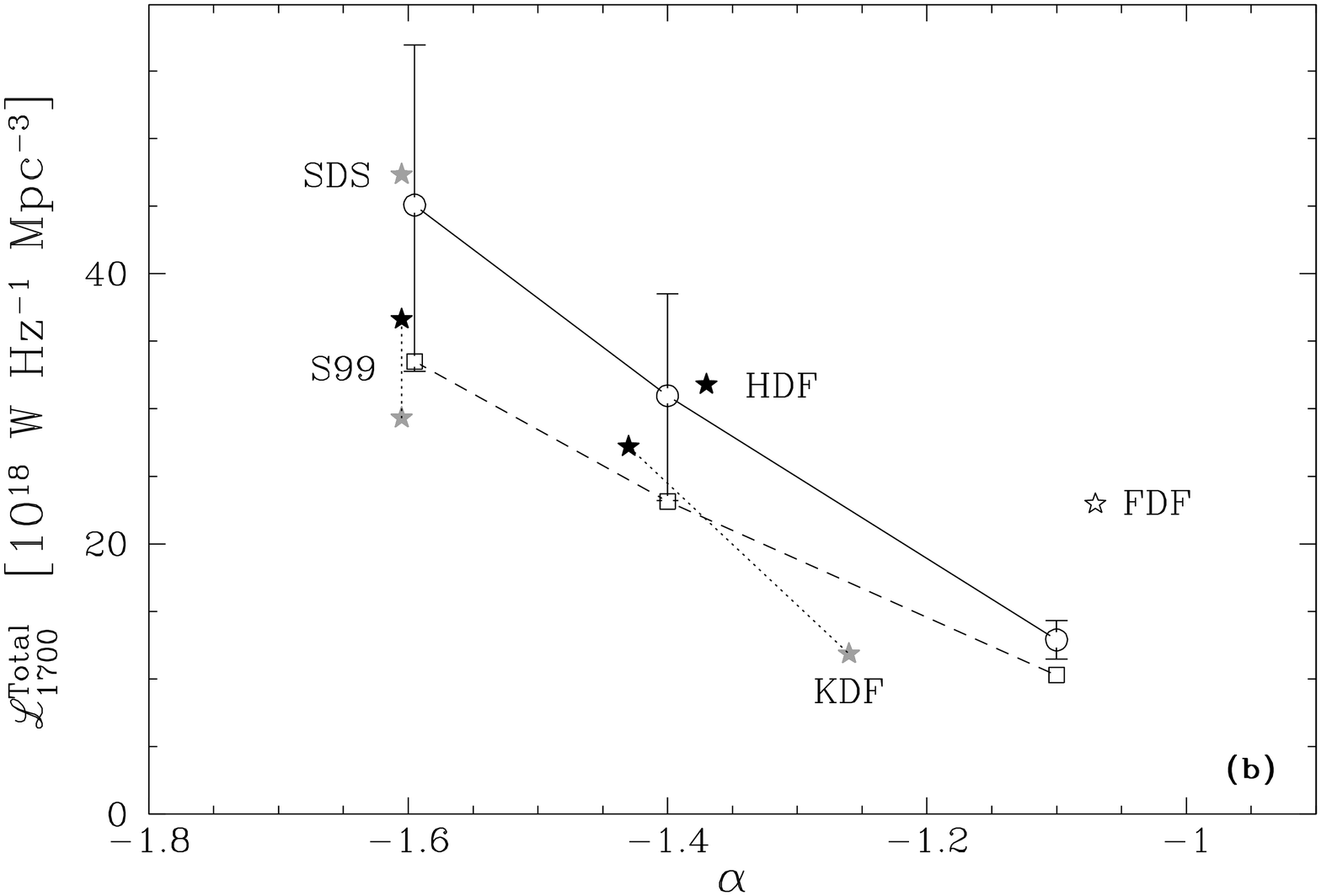}%
  \caption{\label{fig:LD}Luminosity density at $1700$\,\AA\ integrated up to $M=-18.5$ {\bf (a)} and total {\bf (b)} as a function of the assumed value of $\alpha$, based on the photometric-rejection LFs. Circles show the LDs including the flag-0 correction, while squares do not.  Stars show results from other studies (see text). Black, empty and grey stars are for $z\sim 3$, $z\sim 3.5$ and $z\sim 4$ respectively. Symbols at $\alpha=-1.6$ have been slightly shifted for better readability. The different lines connect related symbols for easier identification only.}
\end{figure*}

The major difficulty in the calculation of the luminosity density by integrating the LF is the effect of the $\alpha$ parameter of the Schechter distribution. Indeed, for steep values of $\alpha$, the contribution of the faintest galaxies, where the LF is the least constrained, may dominate that of the much less numerous bright galaxies. The usual work-around is to adopt a limiting absolute magnitude $M_{\mathrm{lim}}$, although its choice is non-physical and rather arbitrary. Even when such choice is warranted, any uncertainty in $\alpha$ may result in significant changes of the LD. We address this problem by calculating the LDs for different fixed values of $\alpha$'s. We follow \citet{SawiThom-2006-LumDep}, who calculated, using the LFs in ST06A, the LD up to $M_{1700}=-18.5$, as well as the total LD. The results are shown in Figs~\ref{fig:LD}a and \ref{fig:LD}b.

Expectedly, the LDs calculated from the uncorrected LFs are systematically lower than those from the corrected ones, the amplitude of this effect being between 20 and 50\%. When integrating up to $M_{1700}=-18.5$, the effect of changing $\alpha$ is moderate, as the three LDs are consistent with a value of ${\cal L}_{1700}(M<-18.5)\simeq 23\,10^{18}$ W Hz$^{-1}$ Mpc$^{-3}$ for the photometric-rejection LFs. This value is about 2.5 times larger than ${\cal L}_{1700}(M<-21.0)$, which is the limit of integration used in S99 to avoid extrapolating the LFs. The density of UV photons from galaxies in the range $-21.0<M_{1700}<-18.5$ is therefore comparable to that of the brightest galaxies with $M_{1700}<-21.0$. When integrating the full LD, the effect of $\alpha$ becomes quite serious, as the LFs differ by about a factor 2 between the $\alpha=-1.1$ and $\alpha=-1.6$ cases. Except for the very steep $\alpha=-1.6$ case, ${\cal L}_{1700}(M<-18.5)$ is $\gtrsim 80$\% of ${\cal L}_{1700}^{\mathrm{Total}}$, meaning that the bulk of the UV photons comes from galaxies brighter than $M^*+3$, i.e.\ 15 times fainter than the characteristic luminosity.

\section{Discussion}

\subsection{Comparison of the LF with previous studies}
\label{sec:comp}
This study of the ultraviolet LF at \deltaz\ is the first one entirely based on a spectroscopic sample. It is therefore important to compare the results in detail with previous works using LBG color selection \citep[S99;][ ST06A]{OuchEtal-2004-CenLym} or photometric redshifts \citep{PoliEtal-2001-EvoLum,GabaEtal-2004-EvoLum}.

With the exception of the KDF sample (ST06A) and the HDF sample \citep{PoliEtal-2001-EvoLum}, some work was needed before we could compare our results with the previous ones, because of differences in the calculation of the LFs. S99 give their STY parameters for a cosmology with $H_0=50$ km s$^{-1}$ Mpc$^{-1}$ and $\Omega_{\mathrm m}=1$; transforming the STY parameters of S99 into the concordance cosmology, we get  $M^*=-21.07$, $\phi^*=1.4\,10^{-3}$ mag$^{-1}$ Mpc$^{-3}$ at $z\sim 3$ and $M^*=-21.12$, $\phi^*=1.07\,10^{-3}$ mag$^{-1}$ Mpc$^{-3}$ at $z\sim 4$. In the case of the FDF \citep{GabaEtal-2004-EvoLum}, a pseudo-filter at 1500\,\AA\ has been used; we therefore corrected their value of $M^*$ by $-0.1$, which corresponds to the difference in $M^*$ we obtain if we calculate the VVDS LFs at 1500\,\AA\ instead of 1700\,\AA. In the case of the SDS \citep{OuchEtal-2004-CenLym}, which found a much steeper $\alpha$ than all other groups, we chose their STY parameters where $\alpha$ has been fixed to $-1.6$.

The above surveys are photometric (however, the S99 sample has been spectroscopically confirmed in \citet{SteiEtal-1996-SpeCon}), which allows them to reach deeper magnitudes than the VVDS, which is a pure magnitude-limited spectroscopic survey. The shallower depth of the VVDS doesn't allow us to constrain the slope $\alpha$ independently. We therefore fixed the value of $\alpha$ to different values found in the literature, and finally adopted $\alpha=-1.4$, compatible with the latest and deepest study of the LF of LBGs (ST06A), as well as with studies based on photometric redshifts in the HDF \citep{PoliEtal-2001-EvoLum,ArnoEtal-2005-MeaEvo}. It cannot be taken for granted, however, that the slope we would have found with a much deeper spectroscopic survey would be compatible with those obtained with Lyman-break or photometric-redshift selections, for instance if these selections miss a significant fraction of the high-redshift galaxies (see Sect.~\ref{sec:newpop}). In any case, the discussion of the value of $\alpha$ is beyond the scope of this paper.

The value we obtain for $\phi^*$ for the PR LF corrected for incompleteness is compatible with those found in studies of the LBG population (S99; ST06A), except for the SDS \citep{OuchEtal-2004-CenLym}, whose LF at $z\sim 4$ is extreme compared to all other studies (see Fig.~\ref{fig:LF}, bottom panels). Photometric-redshift studies of the HDF and FDF find however slightly larger values of $\phi^*$ which are marginally compatible with ours at the 90\% level. We obtained our result by applying only one very simple correction for incompleteness that considers only flag-0 objects. A correction for missed \deltaz\ objects which ended up with flags 1 and 2 in a different redshift range would increase $\phi^*$, but we conclude that we do not find any evidence in the VVDS high-redshift LF for a value of $\phi^*$ in significant disagreement with values found in previous studies.

With a value of $M^*=-21.49$, the characteristic luminosity is found to be $\sim 0.5$\,mag brighter than that found in other studies, which are close to $M^*=-21.0$, or even fainter (SDS). Our $M^*$ estimate differs from these results at the $2.5\sigma$ level at least. We note that $M^*$ may even be brighter if the photometric rejection erroneously discards some very luminous galaxies with correct redshifts (see the values of $M^*$ found for the EW LFs in Table~\ref{tab:lfpar}).

As shown in the bottom panels of Fig.~\ref{fig:LF}, the $M^*$--$\phi^*$ error contours are strongly correlated. If $M^*$ has been underestimated because of statistical fluctuations, $\phi^*$ would also have been underestimated, pushing it to values quite higher than those found in S99 and ST06A at the level of the HDF \citep{PoliEtal-2001-EvoLum} and the FDF \citep{GabaEtal-2004-EvoLum}; however, even pushing the statistics to its limit, it remains difficult to reconcile the characteristic absolute magnitude $M^*$ in the VVDS high-redshift LF with those obtained from other LFs. We discuss this further in the next section.

\subsection{A new population of non-LBG high-redshift galaxies?}
\label{sec:newpop}
Assuming $\alpha=-1.4$, $M^*$ is about $-21.5$, which is $60$\% brighter brighter than all values of $M^*$ obtained so far at this redshift and this wavelength. This quite different $M^*$ needs to be understood.

While Poisson statistics is included in our uncertainties, we checked that cosmic variance on scales commensurable with the size of the VVDS Deep Field is probably negligible compared to our statistical uncertainties. Structures on much larger scales may exist, but are not predicted by theory \citep{SomeEtal-2004-CosVar}. Cosmic variance affecting the fields used by other authors cannot probably explain the differences in the LFs. The KDF survey of ST06A covers about a tenth of the area of the VVDS Deep Field (169\,arcmin$^2$ compared to 1\,720\,arcmin$^2$) split into five fields. The effect of the cosmic variance on $\phi^*$ was found to be of the order of 50\%.

The reasonably good agreement between most previous studies is expected, since they are based on similar observing strategies, and therefore may be subject to similar biases. S99 and ST06A used the same LBG-selection strategy to identify Lyman-break galaxies, clearly different from our pure magnitude-limited selection; SDS \citep{OuchEtal-2004-CenLym} also uses an LBG selection, but with different filters. Studies based on photometric redshifts \citep{PoliEtal-2001-EvoLum,GabaEtal-2004-EvoLum,ArnoEtal-2005-MeaEvo} heavily rely on similar assumptions on galaxy emission and intervening absorption. It is possible that these assumptions do not apply to the entire high-redshift galaxy population. Furthermore, the presence of a strong break in the bluest band remains the most important criterion in identifying a galaxy as a high-redshift object, making the photometric-redshift population little different from that of LBGs in practice. It must be pointed out that straight galaxy counts are much larger at all magnitudes in the VVDS than, for instance, in S99 \citep{LefeEtal-2005-LarPop}, and that the relatively good agreement of $\phi^*$ with LBG-based studies is due to the fact that the latter include substantial incompleteness corrections based on complex simulations that are completely avoided in the VVDS.

\begin{figure}
  \centering
  \includegraphics[width=8.8cm]{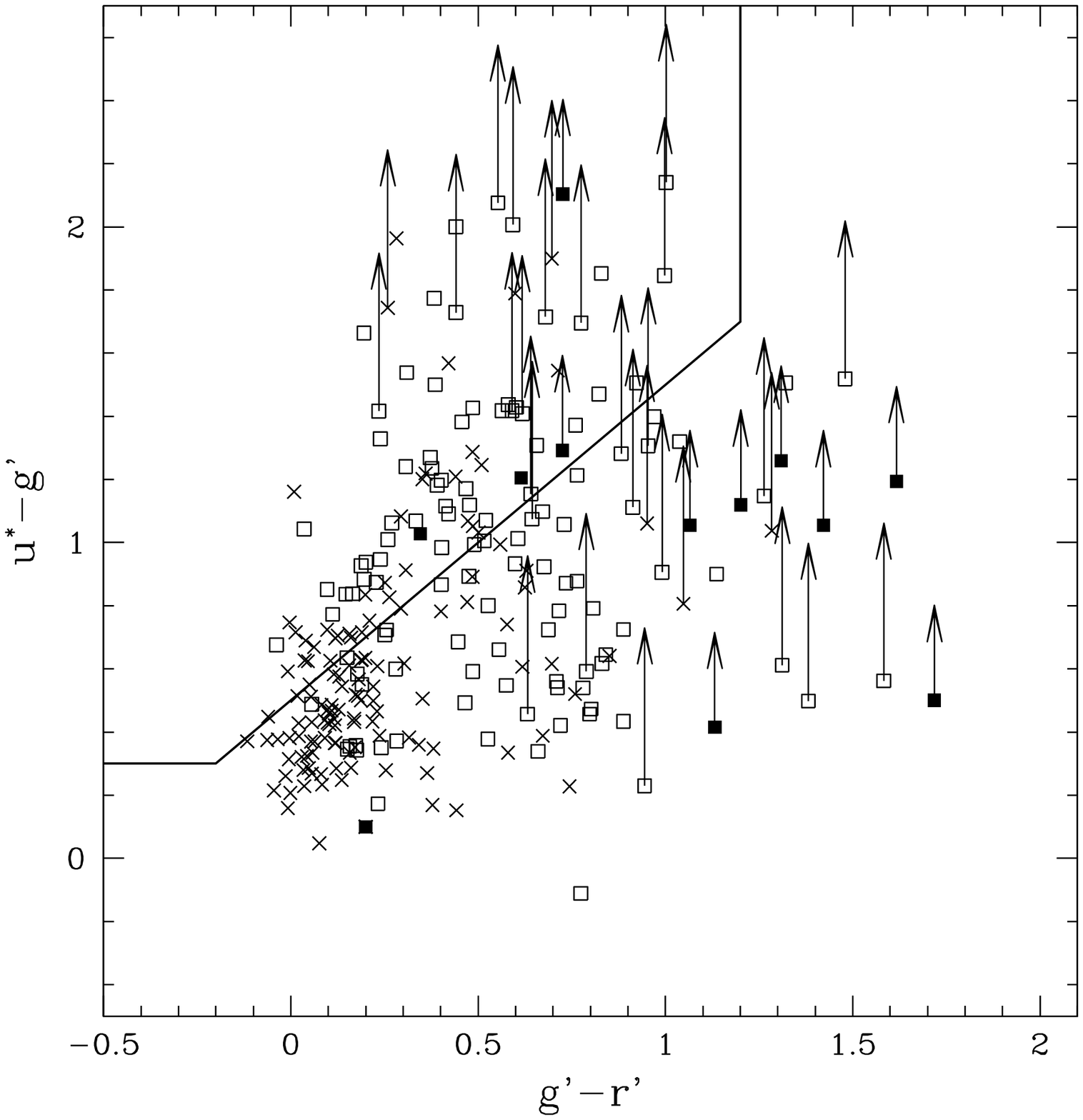}
  \caption{\label{fig:uggr}$(u^*-g')$ vs $(g'-r')$ color-color diagram for the VVDS galaxies with redshifts in the range \deltaz. Upper limits on $u^*$ are indicated by a vertical arrow. Black squares are objects with secure redshifts (flags 3 and 4). Empty squares and crosses are objects with less secure redshifts (flags 1 and 2). A square indicates an object kept in the calculation of the PR LF. The solid line shows the contour of the Lyman-break selection box.}
\end{figure}
\begin{figure}
  \centering
  \includegraphics[width=8.8cm]{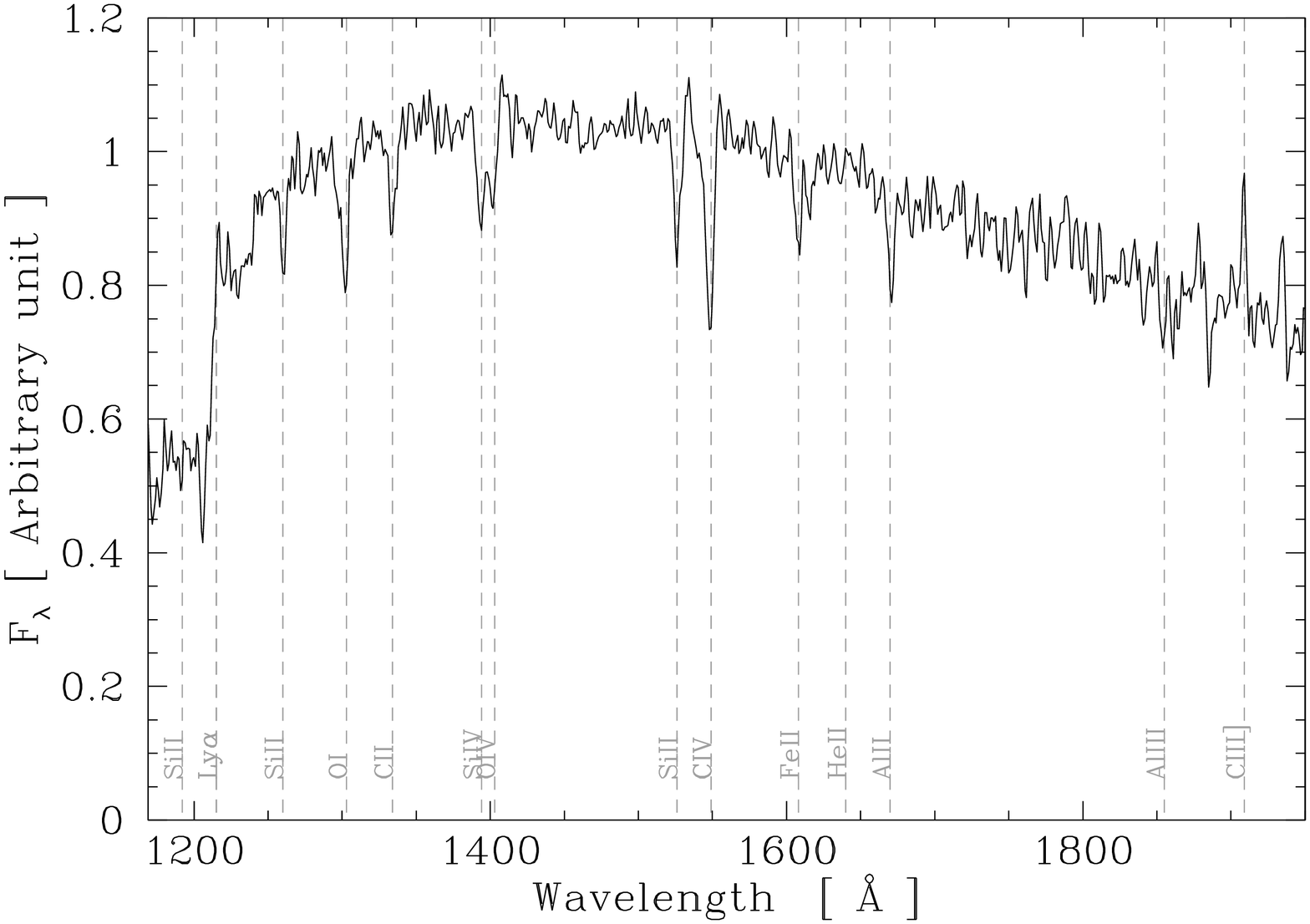}
  \caption{\label{fig:compoout}Composite spectrum of all 172 \deltaz\ galaxy candidates which fall outside of the LBG selection box.}
\end{figure}

The VVDS is essentially free of any assumption on the presence of a break in the continuum; the photometric-rejection LFs \emph{reject} galaxies with very improbable colors, which is not at all the same as \emph{selecting} galaxies based on their colors. The bright $M^*$ found in our survey may therefore be an indication that there exists a population of high-redshift galaxies with ``unexpected'' colors. Fig.~\ref{fig:uggr} shows the $(u^*-g')$ vs $(g'-r')$ color-color diagram for our \deltaz\ galaxies. The Lyman-break selection box has been calculated on the basis of the criteria of \citet{SteiEtal-1996-SpeCon} and adapted to the CFHTLS filters, so that it selects efficiently galaxies with redshifts \deltaz; we have adopted a quite conservative bottom limit for the box, so that it should contain the majority of the LBGs, at the price of some contamination by low-redshift galaxies. Among the twelve secure-redshift galaxies, only four of them fall inside the selection box; three more objects, having only upper limits on $u^*$, may also be inside the box. Five galaxies are however clearly outside the box. Incidentally, Fig.~\ref{fig:uggr} shows that most of the objects rejected for the calculation of the PR LF have $g'-r'<0.4$, a region where numerous low-redshift galaxies are expected, which confirms the validity of the method. Fig.~\ref{fig:compoout} shows the composite spectrum of the 172 galaxies in our \deltaz\ sample whose colors place them outside of the LBG selection box.  The composite spectrum presents very clearly all the expected features of galaxies at $z\sim 3\!-\!4$, both in the spectral lines and in the continuum, making the case for the existence of numerous high-redshift galaxies outside of the selection box quite convincing. One would expect that the LF of VVDS galaxies inside the LBG selection box would match those from S99 and ST06A; however, the calculation of such LFs is quite complex, because it requires the understanding and the modelling of all effects that may result in a failure in the magnitude determination in any of the three filters, an accurate estimate of the contamination by low-redshift galaxies, as well as the ``natural'' crossing of the box boundaries because of the statistical uncertainties in the magnitudes. Such work is therefore outside of the scope of this paper.

One possible reason for the presence of galaxies outside of the LBG selection would be that the intergalactic absorption has been overestimated. Indeed, by recalculating the transmission through the intergalactic medium first derived by \citet{Mada-1995-RadTra}, \citet{Meik-2006-ColCor} found differences in colors of about 0.5--1\,mag for bands containing rest-frame Lyman\,$\alpha$ and shorter wavelengths. Should the bottom of the selection box be lowered by about 0.5\,mag, as suggested by figure 2b in \citet{Meik-2006-ColCor}, it would become hard, if not impossible, to use LBG selections without spectroscopic follow-ups to build complete samples of high-redshift galaxies, as contamination by low-redshift galaxies will be important in the lower part of the box. Particularly strong absorption in the $G$ filter, which is affected by the presence of the Lyman\,$\alpha$ forest, might also contribute to the presence of high-redshift galaxies outside of the LBG selection box.

At redshift $z\sim 2$, it is known that BM/BX-selected galaxies \citep[which is an extension at lower redshift of the LBG color criteria;][]{AdelEtal-2004-OptSel} do not form the complete population of star-forming galaxies. Combining different photometric selection criteria, \citet{ReddEtal-2005-CenOpt} found that the star-forming \emph{BzK} galaxies \citep{DaddEtal-2004-NewPho} do not all satisfy the LBG color criteria, the overlap being only about 70-80\%. The \emph{BzK} color criterion is more efficient at selecting IR-bright galaxies, and these galaxies might have redder $G-{\cal R}$ colors than the BM/BX-selected ones. The VVDS might therefore observe the equivalent of the \emph{BzK} star-forming galaxies at redshifts \deltaz, in addition to the ``normal'' LBG galaxies.

\subsection{The ultraviolet luminosity density in the VVDS}

\begin{figure}
  \centering
  \includegraphics[width=8.8cm]{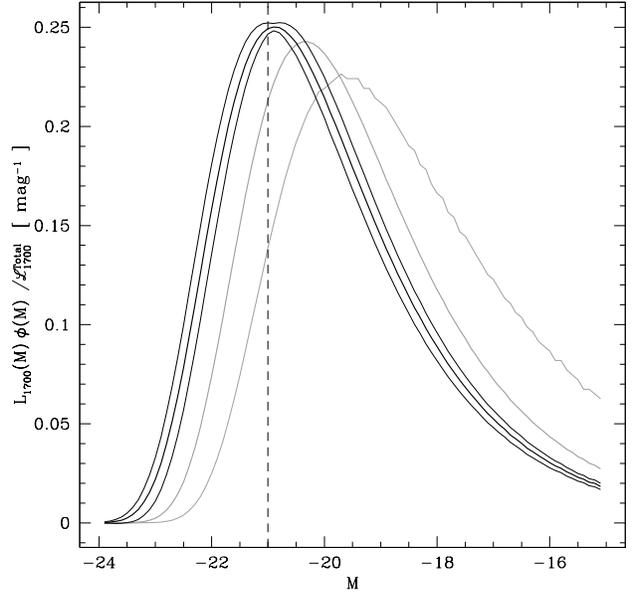}
  \caption{\label{fig:ILD}$L_{1700}(M)\,\phi(M)/{\cal L}_{1700}^{\mathrm{Total}}$ as a function of $M$ for our best estimate of the LD (black line) calculated from the PR LF, compared to the LD at $z\sim 3$ from the KDF (heavy grey line) and at $z\sim 4$ from the SDS (thin grey line). The black thin lines show the 68\% uncertainties on $L_{1700}(M)\,\phi(M)/{\cal L}_{1700}^{\mathrm{Total}}$. The dashed vertical line indicates $M=-21.0$.}
\end{figure}

We compare the UV LD obtained from the VVDS LF at redshifts \deltaz\ with those obtained in the studies discussed in Sect.~\ref{sec:comp}. Most of these studies quote the values of the LD, but, when needed, we calculated the LDs ourselves through the integration of the Schechter distributions they obtained.

In the calculation of the LD, the choice of $\alpha$ plays a very important role, making comparisons between LFs with different $\alpha$ quite meaningless. This effect is quite visible in Figs~\ref{fig:LD}a and (especially) \ref{fig:LD}b. Because of the large error bars inherent in the determination of the LDs, ${\cal L}_{1700}(M<-18.5)$, i.e.\ the LD integrated up to $M_{1700}=-18.5$, and ${\cal L}_{1700}^{\mathrm{Total}}$, the total LD, are mostly compatible with the results from previous works; only the LD at $z\sim 4$ from the KDF \citep{SawiThom-2006-LumDep} being markedly lower.

While the total LD from the VVDS is not very different from the already published values, the significantly brighter $M^*$ we have found modifies strongly the picture of the distribution of the LD as a function of the galaxy luminosity. The VVDS LD integrated up to $M_{1700}=-21.0$ is indeed more than twice that from the KDF at $z\sim 3$ \citep{SawiThom-2006-LumDep}; this is already seen in Fig.~\ref{fig:LF}b, where our non-parametric LF is always at least twice as large as that of ST06A. Galaxies brighter than $M_{1700}=-21.0$ account for a third of the total LD in the VVDS, while this figure is only 15\% in the KDF. The same applies if we compare with other studies \citep[S99;][]{PoliEtal-2001-EvoLum,OuchEtal-2004-CenLym,GabaEtal-2004-EvoLum}. Adopting compatible values for $\alpha$, the ratio ${\cal L}_{1700}(M<-21.0)/{\cal L}_{1700}^{\mathrm{Total}}$ is typically twice as large in the VVDS compared to the other studies. We conclude that the VVDS shows evidence of a more important contribution of the most UV-bright galaxies to the total LD. This is illustrated on Fig.~\ref{fig:ILD}, which shows $L_{1700}(M)\,\phi(M)/{\cal L}_{1700}^{\mathrm{Total}}$, i.e.\ the fraction of the luminosity density at magnitude $M$ to the total LD (see Eq.~(\ref{eq:ld})), as a function of absolute magnitude $M$ for our study and for the KDF study. Comparisons with other studies qualitatively agree, the amplitude of the effect being the strongest for the SDS.

\section{Conclusion}

The UV 1700\,\AA\ luminosity function in the VVDS for galaxies with redshifts \deltaz\ differs significantly from other estimates of the UV LF in the value of the characteristic luminosity, above which the number density of galaxies decreases exponentially; this characteristic luminosity, corresponding to an absolute magnitude $M_{1700}\sim -21.5$, is found to be about 60\% larger than previously reported. On the other hand, the galaxy number density at lower luminosities is mostly compatible with previous studies. Our result has been obtained with minimal corrections for incompleteness, and it is possible that the remaining incompleteness would significantly increase the total luminosity density.

The difference in the LF seems to be due to the fact that Lyman-break selection techniques and photometric-redshift studies are not able to retrieve the full population of high-redshift galaxies. This may be due to a combination of different effects, like, for example, photometric problems that are not well modelled, intergalactic absorption that pushes the bottom of the Lyman-break selection box down in the color-color diagram, or a population of high-redshift galaxies whose colors do not correspond to the expectations of Lyman-break galaxies or photometric redshifts. This would make color-selection techniques unable to build samples of high-redshift galaxies representative of the complete galaxy population. The population of high-redshift luminous galaxies may include a mix of LBGs, the high-redshift equivalent of the \emph{BzK} star-forming galaxy population, and possibly other types of galaxies.

The determination of the luminosity density requires a precise determination of the slope $\alpha$ of the luminosity function at faint magnitudes. Even though some very deep photometric studies like the KDF (ST06A) have put strong constraints on $\alpha$, it is not clear whether the same $\alpha$ would be found if the full population of high-redshift galaxies could be included. In absence of color criteria able to recover the entire population, the determination of $\alpha$ will have to await spectroscopic surveys similar to the VVDS but reaching significantly fainter magnitudes.

While we find that the total luminosity density in the VVDS is compatible with previous estimates, the distribution of the LD among galaxies of different luminosity is quite different, with galaxies brighter than $M_{1700}\sim -21.0$ accounting for about a third of the total LD, a fraction twice as large as previously estimated. Therefore, the VVDS paints a quite different picture of the role of the most actively star-forming galaxies in the history of star formation.

\begin{acknowledgements}
  This research program has been developed within the framework of the VVDS consortium.\\
  This work has been partially supported by the CNRS-INSU and its Programme National de Cosmologie (France), and by Italian Ministry (MIUR) grants COFIN2000 (MM02037133) and COFIN2003 (num.2003020150).\\
  The VLT-VIMOS observations have been carried out on guaranteed time (GTO) allocated by the European Southern Observatory (ESO) to the VIRMOS consortium, under a contractual agreement between the Centre National de la Recherche Scientifique of France, heading a consortium of French and Italian institutes, and ESO, to design, manufacture and test the VIMOS instrument.
\end{acknowledgements}

\bibliographystyle{apj}
\bibliography{biblio}

\begin{thebibliography}{42}
\expandafter\ifx\csname natexlab\endcsname\relax\def\natexlab#1{#1}\fi

\bibitem[{{Abraham} {et~al.}(2004){Abraham}, {Glazebrook}, {McCarthy},
  {Crampton}, {Murowinski}, {J{\o}rgensen}, {Roth}, {Hook}, {Savaglio}, {Chen},
  {Marzke}, \& {Carlberg}}]{AbraEtal-2004-GemDee}
{Abraham}, R.~G., {Glazebrook}, K., {McCarthy}, P.~J., et al. 2004, \aj, 127, 2455

\bibitem[{{Adelberger} {et~al.}(2004){Adelberger}, {Steidel}, {Shapley},
  {Hunt}, {Erb}, {Reddy}, \& {Pettini}}]{AdelEtal-2004-OptSel}
{Adelberger}, K.~L., {Steidel}, C.~C., {Shapley}, A.~E., et al. 2004, \apj, 607, 226

\bibitem[{{Arnouts} {et~al.}(2002){Arnouts}, {Moscardini}, {Vanzella},
  {Colombi}, {Cristiani}, {Fontana}, {Giallongo}, {Matarrese}, \&
  {Saracco}}]{ArnoEtal-2002-MeaRed}
{Arnouts}, S., {Moscardini}, L., {Vanzella}, E., et al. 2002,
  \mnras, 329, 355

\bibitem[{{Arnouts} {et~al.}(2005){Arnouts}, {Schiminovich}, {Ilbert},
  {Tresse}, {Milliard}, {Treyer}, {Bardelli}, {Budavari}, {Wyder}, {Zucca}, {Le
  F{\`e}vre}, {Martin}, {Vettolani}, {Adami}, {Arnaboldi}, {Barlow}, {Bianchi},
  {Bolzonella}, {Bottini}, {Byun}, {Cappi}, {Charlot}, {Contini}, {Donas},
  {Forster}, {Foucaud}, {Franzetti}, {Friedman}, {Garilli}, {Gavignaud},
  {Guzzo}, {Heckman}, {Hoopes}, {Iovino}, {Jelinsky}, {Le Brun}, {Lee},
  {Maccagni}, {Madore}, {Malina}, {Marano}, {Marinoni}, {McCracken}, {Mazure},
  {Meneux}, {Merighi}, {Morrissey}, {Neff}, {Paltani}, {Pell{\`o}}, {Picat},
  {Pollo}, {Pozzetti}, {Radovich}, {Rich}, {Scaramella}, {Scodeggio},
  {Seibert}, {Siegmund}, {Small}, {Szalay}, {Welsh}, {Xu}, {Zamorani}, \&
  {Zanichelli}}]{ArnoEtal-2005-MeaEvo}
{Arnouts}, S., {Schiminovich}, D., {Ilbert}, et al. 2005, \apjl, 619, L43

\bibitem[{{Bottini} {et~al.}(2005){Bottini}, {Garilli}, {Maccagni}, {Tresse},
  {Le Brun}, {Le F{\`e}vre}, {Picat}, {Scaramella}, {Scodeggio}, {Vettolani},
  {Zanichelli}, {Adami}, {Arnaboldi}, {Arnouts}, {Bardelli}, {Bolzonella},
  {Cappi}, {Charlot}, {Ciliegi}, {Contini}, {Foucaud}, {Franzetti}, {Guzzo},
  {Ilbert}, {Iovino}, {McCracken}, {Marano}, {Marinoni}, {Mathez}, {Mazure},
  {Meneux}, {Merighi}, {Paltani}, {Pollo}, {Pozzetti}, {Radovich}, {Zamorani},
  \& {Zucca}}]{BottEtal-2005-Vmmps}
{Bottini}, D., {Garilli}, B., {Maccagni}, D., et al. 2005, \pasp,
  117, 996

\bibitem[{{Chapman} {et~al.}(2005){Chapman}, {Blain}, {Smail}, \&
  {Ivison}}]{ChapEtal-2005-RedSur}
{Chapman}, S.~C., {Blain}, A.~W., {Smail}, I., \& {Ivison}, R.~J. 2005, \apj,
  622, 772

\bibitem[{{Coleman} {et~al.}(1980){Coleman}, {Wu}, \&
  {Weedman}}]{ColeEtal-1980-ColMag}
{Coleman}, G.~D., {Wu}, C.-C., \& {Weedman}, D.~W. 1980, \apjs, 43, 393

\bibitem[{{Daddi} {et~al.}(2004){Daddi}, {Cimatti}, {Renzini}, {Fontana},
  {Mignoli}, {Pozzetti}, {Tozzi}, \& {Zamorani}}]{DaddEtal-2004-NewPho}
{Daddi}, E., {Cimatti}, A., {Renzini}, A., et al. 2004, \apj, 617, 746

\bibitem[{{Efstathiou} {et~al.}(1988){Efstathiou}, {Ellis}, \&
  {Peterson}}]{EfstEtal-1988-AnaCom}
{Efstathiou}, G., {Ellis}, R.~S., \& {Peterson}, B.~A. 1988, \mnras, 232, 431

\bibitem[{{Fioc} \& {Rocca-Volmerange}(1997)}]{FiocRocc-1997-PegUV}
{Fioc}, M. \& {Rocca-Volmerange}, B. 1997, \aap, 326, 950

\bibitem[{{Fontana} {et~al.}(2000){Fontana}, {D'Odorico}, {Poli}, {Giallongo},
  {Arnouts}, {Cristiani}, {Moorwood}, \& {Saracco}}]{FontaEtal-2000-PhoRed}
{Fontana}, A., {D'Odorico}, S., {Poli}, F., et al. 2000, \aj, 120, 2206

\bibitem[{{Gabasch} {et~al.}(2004){Gabasch}, {Bender}, {Seitz}, {Hopp},
  {Saglia}, {Feulner}, {Snigula}, {Drory}, {Appenzeller}, {Heidt}, {Mehlert},
  {Noll}, {B{\"o}hm}, {J{\"a}ger}, {Ziegler}, \&
  {Fricke}}]{GabaEtal-2004-EvoLum}
{Gabasch}, A., {Bender}, R., {Seitz}, S., et al. 2004, \aap, 421, 41

\bibitem[{{Ilbert} {et~al.}(2006{\natexlab{a}}){Ilbert}, {Arnouts},
  {McCracken}, {Bolzonella}, {Bertin}, {Le F{\`e}vre}, {Mellier}, {Zamorani},
  {Pell{\`o}}, {Iovino}, {Tresse}, {Le Brun}, {Bottini}, {Garilli}, {Maccagni},
  {Picat}, {Scaramella}, {Scodeggio}, {Vettolani}, {Zanichelli}, {Adami},
  {Bardelli}, {Cappi}, {Charlot}, {Ciliegi}, {Contini}, {Cucciati}, {Foucaud},
  {Franzetti}, {Gavignaud}, {Guzzo}, {Marano}, {Marinoni}, {Mazure}, {Meneux},
  {Merighi}, {Paltani}, {Pollo}, {Pozzetti}, {Radovich}, {Zucca}, {Bondi},
  {Bongiorno}, {Busarello}, {de La Torre}, {Gregorini}, {Lamareille}, {Mathez},
  {Merluzzi}, {Ripepi}, {Rizzo}, \& {Vergani}}]{IlbeEtal-2006-AccPho}
{Ilbert}, O., {Arnouts}, S., {McCracken}, H.~J., et al.
  2006{\natexlab{a}}, \aap, 457, 841

\bibitem[{{Ilbert} {et~al.}(2006{\natexlab{b}}){Ilbert}, {Cucciati},
  {Marinoni}, {Tresse}, {Le Fevre}, {Zamorani}, {Bardelli}, {Iovino}, {Zucca},
  {Arnouts}, {Bottini}, {Garilli}, {Le Brun}, {Maccagni}, {Picat},
  {Scaramella}, {Scodeggio}, {Vettolani}, {Zanichelli}, {Adami}, {Bolzonella},
  {Cappi}, {Charlot}, {Ciliegi}, {Contini}, {Foucaud}, {Franzetti},
  {Gavignaud}, {Guzzo}, {Marano}, {Mazure}, {McCracken}, {Meneux}, {Merighi},
  {Paltani}, {Pello}, {Pollo}, {Pozzetti}, {Radovich}, {Bondi}, {Bongiorno},
  {Busarello}, {De La Torre}, {Gregorini}, {Lamareille}, {Mathez}, {Mellier},
  {Merluzzi}, {Ripepi}, {Rizzo}, \& {Vergani}}]{IlbeEtal-2006-EviEnv}
{Ilbert}, O., {Cucciati}, O., {Marinoni}, C., et al. 2006{\natexlab{b}}, A\&A in press. ArXiv
  Astrophysics e-prints: astro-ph/0602329

\bibitem[{{Ilbert} {et~al.}(2006{\natexlab{c}}){Ilbert}, {Lauger}, {Tresse},
  {Buat}, {Arnouts}, {Le F{\`e}vre}, {Burgarella}, {Zucca}, {Bardelli},
  {Zamorani}, {Bottini}, {Garilli}, {Le Brun}, {Maccagni}, {Picat},
  {Scaramella}, {Scodeggio}, {Vettolani}, {Zanichelli}, {Adami}, {Arnaboldi},
  {Bolzonella}, {Cappi}, {Charlot}, {Contini}, {Foucaud}, {Franzetti},
  {Gavignaud}, {Guzzo}, {Iovino}, {McCracken}, {Marano}, {Marinoni}, {Mathez},
  {Mazure}, {Meneux}, {Merighi}, {Paltani}, {Pello}, {Pollo}, {Pozzetti},
  {Radovich}, {Bondi}, {Bongiorno}, {Busarello}, {Ciliegi}, {Mellier},
  {Merluzzi}, {Ripepi}, \& {Rizzo}}]{IlbeEtal-2006-GalLum}
{Ilbert}, O., {Lauger}, S., {Tresse}, L., et al. 2006{\natexlab{c}}, \aap, 453, 809

\bibitem[{{Ilbert} {et~al.}(2004){Ilbert}, {Tresse}, {Arnouts}, {Zucca},
  {Bardelli}, {Zamorani}, {Adami}, {Cappi}, {Garilli}, {Le F{\`e}vre},
  {Maccagni}, {Meneux}, {Scaramella}, {Scodeggio}, {Vettolani}, \&
  {Zanichelli}}]{IlbeEtal-2004-BiaEst}
{Ilbert}, O., {Tresse}, L., {Arnouts}, S., et al. 2004, \mnras, 351, 541

\bibitem[{{Ilbert} {et~al.}(2005){Ilbert}, {Tresse}, {Zucca}, {Bardelli},
  {Arnouts}, {Zamorani}, {Pozzetti}, {Bottini}, {Garilli}, {Le Brun}, {Le
  F{\`e}vre}, {Maccagni}, {Picat}, {Scaramella}, {Scodeggio}, {Vettolani},
  {Zanichelli}, {Adami}, {Arnaboldi}, {Bolzonella}, {Cappi}, {Charlot},
  {Contini}, {Foucaud}, {Franzetti}, {Gavignaud}, {Guzzo}, {Iovino},
  {McCracken}, {Marano}, {Marinoni}, {Mathez}, {Mazure}, {Meneux}, {Merighi},
  {Paltani}, {Pello}, {Pollo}, {Radovich}, {Bondi}, {Bongiorno}, {Busarello},
  {Ciliegi}, {Lamareille}, {Mellier}, {Merluzzi}, {Ripepi}, \&
  {Rizzo}}]{IlbeEtal-2005-EvoGal}
{Ilbert}, O., {Tresse}, L., {Zucca}, E., et al. 2005, \aap,
  439, 863

\bibitem[{{Le F{\` e}vre} {et~al.}(2004){Le F{\` e}vre}, {Mellier},
  {McCracken}, {Foucaud}, {Gwyn}, {Radovich}, {Dantel-Fort}, {Bertin},
  {Moreau}, {Cuillandre}, {Pierre}, {Le Brun}, {Mazure}, \&
  {Tresse}}]{LefeEtal-2004-VdiI}
{Le F{\` e}vre}, O., {Mellier}, Y., {McCracken}, H.~J., et al. 2004, \aap, 417, 839

\bibitem[{{Le F{\`e}vre} {et~al.}(2005{\natexlab{a}}){Le F{\`e}vre}, {Paltani},
  {Arnouts}, {Charlot}, {Foucaud}, {Ilbert}, {McCracken}, {Zamorani},
  {Bottini}, {Garilli}, {Le Brun}, {Maccagni}, {Picat}, {Scaramella},
  {Scodeggio}, {Tresse}, {Vettolani}, {Zanichelli}, {Adami}, {Bardelli},
  {Bolzonella}, {Cappi}, {Ciliegi}, {Contini}, {Franzetti}, {Gavignaud},
  {Guzzo}, {Iovino}, {Marano}, {Marinoni}, {Mazure}, {Meneux}, {Merighi},
  {Pell{\`o}}, {Pollo}, {Pozzetti}, {Radovich}, {Zucca}, {Arnaboldi}, {Bondi},
  {Bongiorno}, {Busarello}, {Gregorini}, {Lamareille}, {Mathez}, {Mellier},
  {Merluzzi}, {Ripepi}, \& {Rizzo}}]{LefeEtal-2005-LarPop}
{Le F{\`e}vre}, O., {Paltani}, S., {Arnouts}, S., et al. 2005{\natexlab{a}}, \nat, 437, 519

\bibitem[{{Le F{\`e}vre} {et~al.}(2005{\natexlab{b}}){Le F{\`e}vre},
  {Vettolani}, {Garilli}, {Tresse}, {Bottini}, {Le Brun}, {Maccagni}, {Picat},
  {Scaramella}, {Scodeggio}, {Zanichelli}, {Adami}, {Arnaboldi}, {Arnouts},
  {Bardelli}, {Bolzonella}, {Cappi}, {Charlot}, {Ciliegi}, {Contini},
  {Foucaud}, {Franzetti}, {Gavignaud}, {Guzzo}, {Ilbert}, {Iovino},
  {McCracken}, {Marano}, {Marinoni}, {Mathez}, {Mazure}, {Meneux}, {Merighi},
  {Paltani}, {Pell{\`o}}, {Pollo}, {Pozzetti}, {Radovich}, {Zamorani}, {Zucca},
  {Bondi}, {Bongiorno}, {Busarello}, {Lamareille}, {Mellier}, {Merluzzi},
  {Ripepi}, \& {Rizzo}}]{LefeEtal-2005-FirEpo}
{Le F{\`e}vre}, O., {Vettolani}, G., {Garilli}, B., et al.
  2005{\natexlab{b}}, \aap, 439, 845

\bibitem[{Le~F\`evre {et~al.}(2003)Le~F\`evre, Vettolani, Maccagni, Mancini,
  Mazure, Mellier, Picat, Arnaboldi, Bardelli, Bertin, Busarello, Cappi,
  Charlot, Chincarini, Colombi, Dantel-Fort, Foucaud, Garilli, Guzzo, Iovino,
  Marinoni, Mathez, McCracken, Pello, Radovich, Ripepi, Saracco, Scaramella,
  Scoreggio, Tresse, Zanichelli, Zamorani, \& Zucca}]{LefeEtal-2003-VirVLT}
Le~F\`evre, O., Vettolani, G., Maccagni, D., et al.
  2003. In: Discoveries and Research Prospects from 6- to 10-Meter-Class
  Telescopes II. P. Guhathakurta (Ed.). Proceedings of the SPIE,
  4834, 173

\bibitem[{{Madau}(1995)}]{Mada-1995-RadTra}
{Madau}, P. 1995, \apj, 441, 18

\bibitem[{{McCracken} {et~al.}(2003){McCracken}, {Radovich}, {Bertin},
  {Mellier}, {Dantel-Fort}, {Le F{\` e}vre}, {Cuillandre}, {Gwyn}, {Foucaud},
  \& {Zamorani}}]{MccrEtal-2003-VdiII}
{McCracken}, H.~J., {Radovich}, M., {Bertin}, E., et al. 2003, \aap, 410, 17

\bibitem[{{Meiksin}(2006)}]{Meik-2006-ColCor}
{Meiksin}, A. 2006, \mnras, 365, 807

\bibitem[{{Ouchi} {et~al.}(2004){Ouchi}, {Shimasaku}, {Okamura}, {Furusawa},
  {Kashikawa}, {Ota}, {Doi}, {Hamabe}, {Kimura}, {Komiyama}, {Miyazaki},
  {Miyazaki}, {Nakata}, {Sekiguchi}, {Yagi}, \&
  {Yasuda}}]{OuchEtal-2004-CenLym}
{Ouchi}, M., {Shimasaku}, K., {Okamura}, S., et al. 2004, \apj, 611, 660

\bibitem[{{Poli} {et~al.}(2001){Poli}, {Menci}, {Giallongo}, {Fontana},
  {Cristiani}, \& {D'Odorico}}]{PoliEtal-2001-EvoLum}
{Poli}, F., {Menci}, N., {Giallongo}, E., et al. 2001, \apjl, 551, L45

\bibitem[{{Reddy} {et~al.}(2005){Reddy}, {Erb}, {Steidel}, {Shapley},
  {Adelberger}, \& {Pettini}}]{ReddEtal-2005-CenOpt}
{Reddy}, N.~A., {Erb}, D.~K., {Steidel}, C.~C., et al. 2005, \apj, 633, 748

\bibitem[{{Rowan-Robinson}(2003)}]{Rowa-2003-PhoRed}
{Rowan-Robinson}, M. 2003, \mnras, 345, 819

\bibitem[{{Sandage} {et~al.}(1979){Sandage}, {Tammann}, \&
  {Yahil}}]{SandEtal-1979-VelFie}
{Sandage}, A., {Tammann}, G.~A., \& {Yahil}, A. 1979, \apj, 232, 352

\bibitem[{{Saracco} {et~al.}(2006){Saracco}, {Fiano}, {Chincarini}, {Vanzella},
  {Longhetti}, {Cristiani}, {Fontana}, {Giallongo}, \&
  {Nonino}}]{SaraEtal-2006-ProEvo}
{Saracco}, P., {Fiano}, A., {Chincarini}, G., et al. 2006,
  \mnras, 367, 349

\bibitem[{{Sawicki} \& {Thompson}(2006{\natexlab{a}})}]{SawiThom-2005-UVGal}
{Sawicki}, M. \& {Thompson}, D. 2006{\natexlab{a}}, \apj, 642, 653 -- ST06A

\bibitem[{{Sawicki} \& {Thompson}(2006{\natexlab{b}})}]{SawiThom-2006-LumDep}
{Sawicki}, M. \& {Thompson}, D. 2006{\natexlab{b}}, \apj, 648, 299

\bibitem[{{Schechter}(1976)}]{Sche-1976-AnaExp}
{Schechter}, P. 1976, \apj, 203, 297

\bibitem[{{Schmidt}(1968)}]{Schm-1968-SpaDis}
{Schmidt}, M. 1968, \apj, 151, 393

\bibitem[{{Scodeggio} {et~al.}(2005){Scodeggio}, {Franzetti}, {Garilli},
  {Zanichelli}, {Paltani}, {Maccagni}, {Bottini}, {Le Brun}, {Contini},
  {Scaramella}, {Adami}, {Bardelli}, {Zucca}, {Tresse}, {Ilbert}, {Foucaud},
  {Iovino}, {Merighi}, {Zamorani}, {Gavignaud}, {Rizzo}, {McCracken}, {Le
  F{\`e}vre}, {Picat}, {Vettolani}, {Arnaboldi}, {Arnouts}, {Bolzonella},
  {Cappi}, {Charlot}, {Ciliegi}, {Guzzo}, {Marano}, {Marinoni}, {Mathez},
  {Mazure}, {Meneux}, {Pell{\`o}}, {Pollo}, {Pozzetti}, \&
  {Radovich}}]{ScodEtal-2005-Vipgi}
{Scodeggio}, M., {Franzetti}, P., {Garilli}, B., et al. 2005, \pasp,
  117, 1284

\bibitem[{{Somerville} {et~al.}(2004){Somerville}, {Lee}, {Ferguson},
  {Gardner}, {Moustakas}, \& {Giavalisco}}]{SomeEtal-2004-CosVar}
{Somerville}, R.~S., {Lee}, K., {Ferguson}, H.~C., et al. 2004, \apjl, 600, L171

\bibitem[{{Steidel} {et~al.}(1999){Steidel}, {Adelberger}, {Giavalisco},
  {Dickinson}, \& {Pettini}}]{SteiEtal-1999-LymBre}
{Steidel}, C.~C., {Adelberger}, K.~L., {Giavalisco}, M., {Dickinson}, M., \&
  {Pettini}, M. 1999, \apj, 519, 1 -- S99

\bibitem[{{Steidel} {et~al.}(1996){Steidel}, {Giavalisco}, {Pettini},
  {Dickinson}, \& {Adelberger}}]{SteiEtal-1996-SpeCon}
{Steidel}, C.~C., {Giavalisco}, M., {Pettini}, M., {Dickinson}, M., \&
  {Adelberger}, K.~L. 1996, \apjl, 462, L17

\bibitem[{{Tresse} {et al.}(2006)}]{TresEtal-2006-VVDSLD}
{Tresse}, L., {Ilbert}, O., {Zucca}, E., {et al.} 2006, Submitted to A\&A. ArXiv Astrophysics e-prints: astro-ph/0609005

\bibitem[{{van Dokkum} {et~al.}(2003){van Dokkum}, {F{\"o}rster Schreiber},
  {Franx}, {Daddi}, {Illingworth}, {Labb{\'e}}, {Moorwood}, {Rix},
  {R{\"o}ttgering}, {Rudnick}, {van der Wel}, {van der Werf}, \& {van
  Starkenburg}}]{vanDEtal-2003-SpeCon}
{van Dokkum}, P.~G., {F{\"o}rster Schreiber}, N.~M., {Franx}, M., et al. 2003, \apjl, 587, L83

\bibitem[{{Zucca} {et al.}(2006)}]{ZuccEtal-LFTyp}
{Zucca}, E., {Ilbert}, O., {Bardelli}, S., et al. 2006, \aap, 455, 879

\bibitem[{{Zucca} {et~al.}(1997){Zucca}, {Zamorani}, {Vettolani}, {Cappi},
  {Merighi}, {Mignoli}, {Stirpe}, {MacGillivray}, {Collins}, {Balkowski},
  {Cayatte}, {Maurogordato}, {Proust}, {Chincarini}, {Guzzo}, {Maccagni},
  {Scaramella}, {Blanchard}, \& {Ramella}}]{ZuccEtal-1997-ESOSli}
{Zucca}, E., {Zamorani}, G., {Vettolani}, G., et al. 1997, \aap, 326, 477

\end{thebibliography}

\end{document}